\documentclass[pre,aps,twocolumn,amsmath,amssymb,showpacs]{revtex4-1}
\usepackage{amsmath}
\usepackage{amssymb}
\usepackage{graphicx}
\usepackage{hyperref}
\usepackage{float}
\usepackage{subfigure}
\usepackage[utf8x]{inputenc}
\usepackage{gensymb}
\usepackage{color}

\begin{document}
\title{Fluid-fluid demixing and density anomaly in a ternary mixture of hard spheres}

\author{Nathann T. Rodrigues}\email{nathan.rodrigues@ufv.br}
\affiliation{Departamento de F\'isica, Universidade Federal de Vi\c cosa, 36570-900, Vi\c cosa, MG, Brazil}
\author{Tiago J. Oliveira}\email{tiago@ufv.br}
\affiliation{Departamento de F\'isica, Universidade Federal de Vi\c cosa, 36570-900, Vi\c cosa, MG, Brazil}

\date{\today}

\begin{abstract}
We report the grand-canonical solution of a ternary mixture of discrete hard spheres defined on a Husimi lattice built with cubes, which provides a mean-field approximation for this system on the cubic lattice. The mixture is composed by point-like particles (0NN) and particles which exclude up to their first (1NN) and second neighbors (2NN), with activities $z_0$, $z_1$ and $z_2$, respectively. Our solution reveals a very rich thermodynamic behavior, with two solid phases associated with the ordering of 1NN ($S1$) or 2NN particles ($S2$), and two fluid phases, being one regular ($RF$) and the other characterized by a dominance of 0NN particles ($F0$ phase). However, in most part of the phase diagram these fluid ($F$) phases are indistinguishable. Discontinuous transitions are observed between all the four phases, yielding several coexistence surfaces in the system, among which a fluid-fluid and a solid-solid demixing surface. The former one is limited by a line of critical points and a line of triple points (where the phases $RF$-$F0$-$S2$ coexist), both meeting at a special point, after which the fluid-fluid coexistence becomes metastable. Another line of triple points is found, connecting the $F$-$S1$, $F$-$S2$ and $S1$-$S2$ coexistence surfaces. A critical $F$-$S1$ surface is also observed meeting the $F$-$S1$ coexistence one at a line of tricritical points. Furthermore, a thermodynamic anomaly characterized by minima in isobaric curves of the total density of particles is found, yielding three surfaces of minimal density in the activity space, depending on which activity is kept fixed during its calculation.
\end{abstract}


\maketitle

\section{Introduction}
\label{intro}

In two recent works \cite{NT0NN2NN,NT0NN1NN-1NN2NN}, we have investigated three athermal binary mixtures of hard spheres placed on the cubic lattice, where they are approximated by $k$NN particles, i.e., particles which exclude up their $k$th neighbors. The mixtures considered were composed by pairs of the three smallest ``spheres'' (0NN-1NN, 0NN-2NN and 1NN-2NN) and grand-canonical phase diagrams for these systems were obtained through their solution on a Husimi lattice built with cubes \cite{NT0NN2NN,NT0NN1NN-1NN2NN} [see Fig. \ref{fig1}(a)]. Such mean-field solutions display very rich entropy-driven thermodynamic behaviors. In a brief, in the 0NN-1NN case, a fluid ($F$) and a solid ($S1$) phase (associated with the ordering of 1NN particles) were found, separated by a continuous and a discontinuous transition line, both meeting at a tricritical point \cite{NT0NN1NN-1NN2NN}. The same phases and behavior were observed in the 1NN-2NN mixture, but now another solid ($S2$) phase (associated with the ordering of 2NN particles) is present in the system and it is separated from the $F$ and $S1$ phases by coexistence lines, which meet the $F$-$S1$ one at a triple point \cite{NT0NN1NN-1NN2NN}. For the 0NN-2NN mixture, an even more interesting scenario was found, with two fluid phases, one regular ($RF$) and another characterized by a dominance of 0NN particles ($F0$), beyond the $S2$ phase. Its phase diagram displays a fluid-fluid demixing transition, characterized by a $RF$-$F0$ coexistence line, which ends in a critical point and also meet a $RF$-$S2$ and a $F0$-$S2$ coexistence line in a triple point \cite{NT0NN2NN}. Another interesting feature observed in the 0NN-1NN and 0NN-2NN mixtures is a thermodynamic anomaly characterized by minima in isobaric curves of the total density of particles. These different behaviors in the binary systems --- specially the absence of the density anomaly in the 1NN-2NN case and the existence of the fluid-fluid demixing only the 0NN-2NN mixture --- lead us to inquiry how these things are connected in the more general and interesting case of the 0NN-1NN-2NN ternary mixture. In this paper we address this by tackling the challenging problem of building up the three-dimensional (3D) phase diagram for this ternary model, once again by solving it on a cubic Husimi lattice.

We remark that ternary mixtures are central in a vast number of systems. To name a few, we may cite ternary quasicrystals \cite{QCs}, several chemical reactions involving three species (e.g., 2H$_2$ + O$_2$ $\leftrightharpoons$ 2H$_2$O), mixtures of cholesterol and lipids which play a key role in cell membranes \cite{Elliott,Solanko}, water-oil-surfactant mixtures \cite{Sager,Abbott}, and so on. Moreover, colloids and granular matter in general are usually modeled as hard spheres \cite{granular}. In fact, (thermal) models of ternary mixtures have recently been used to investigate, e.g., critical Casimir forces among colloidal particles dispersed in a binary solvent \cite{Tasios,Maciolek} and the percolation of patchy colloids \cite{Seiferling,Teixeira}. Regarding the athermal case, ternary mixtures of hard spheres have been theoretically investigated using different approximation methods \cite{Gazzillo,Fotouh,Paschinger,Konig} and in several works such approaches were compared with both simulations \cite{Schaink,Santos,Santos2,Malijevsk,Malijevsk2,Patra,Yu} and experiments \cite{Hoshino,Fotouh2,Matsunaga,Boned,Anderson,Yi}. 

Despite these works considering the hard spheres in the continuous space, as long as we know, ternary mixtures of them on the lattice (i.e., of $k$NN particles) have never been investigated so far. Actually, even binary mixtures of $k$NN particles on the cubic (or any other three-dimensional) lattice have never be studied before our recent works discussed above, based on the cubic Husimi lattice solution \cite{NT0NN1NN-1NN2NN,NT0NN2NN}. On the other hand, the 0NN-1NN mixture has been considered by several authors on the triangular \cite{frenkel0nn1nn,Lekkerkerker,Nienhuis} and square lattice \cite{poland,Jim01,tiago15}, beyond a mean-field solution on the Bethe lattice \cite{tiago11}. A central topic in these studies is the possibility of a fluid-fluid demixing transition in $k$NN-binary mixtures. We remark that in nonadditive mixtures, as is the case in $k$NN ones, the demixed phases fill the space more effectively than the mixed one, so that a demixing is expected; and it has indeed been observed in several systems \cite{Dijkstra,Schmidt,Brader,Dickman95,Roiji,Wensink,Dubois,Varga,Mederos,Schmidt02,Heras}. In the $k$NN systems, while  van Duijneveldt and Lekkerkerker \cite{Lekkerkerker} claimed to have found a fluid-fluid demixing in the 0NN-1NN mixture on the triangular lattice, strong evidence against this have been presented by other authors \cite{frenkel0nn1nn,Nienhuis}, who observed only fluid-solid transitions in this model, similarly to what is found for this mixture on the square \cite{poland,Jim01,tiago15} and hierarchical lattices \cite{tiago11,NT0NN1NN-1NN2NN}. 

Therefore, it seems that among all $k$NN-binary mixtures investigated so far the 0NN-2NN case on the cubic Husimi lattice is the single one presenting a stable fluid-fluid demixing \cite{NT0NN2NN}. Hence, it is quite interesting to analyze what happens with this transition when 1NN particles are introduced in this system. As will be demonstrated below, the $RF$-$F0$ coexistence still exists for small densities of 1NN particles, giving rise to a fluid-fluid demixing surface in the 3D phase diagram of the 0NN-1NN-2NN mixture. Such phase diagram is also featured by several other coexistence surfaces, including a solid-solid demixing one, a critical $F$-$S1$ surface and lines of critical, tricritical and triple points, giving rise to a very rich thermodynamic behavior.

The rest of the paper is organized as follows. In Sec. \ref{defmod} the model is defined and solved on the cubic Husimi lattice in terms of recursion relations, which are presented in the appendix. Its thermodynamic behavior is discussed in Sec. \ref{therbe}. Our final discussions and concluding remarks are presented in Sec. \ref{SecConclusions}.

\section{Model definition and solution on the Husimi lattice}
\label{defmod}

The model considered here consists of a ternary mixture of hard spheres, with diameter $\lambda$, defined on --- and centered at the vertices of --- a cubic lattice. By assuming the lattice spacing being $a$, the smallest (0NN) particles corresponds to spheres of diameter $\lambda=a$, so that they occupy a single lattice site and effectively do not interact with each other. The intermediate (1NN) and largest (2NN) particles are spheres of diameters $\lambda=\sqrt{2}a$ and $\lambda=\sqrt{3}a$, respectively, which thus exclude up to their nearest and next-nearest neighbors of being occupied by other particles. Let us remark that in our previous works on binary mixtures \cite{NT0NN1NN-1NN2NN,NT0NN2NN}, we have wrongly stated that the 1NN and 2NN particles would correspond to \textit{cubes} of lateral sizes $\lambda=\sqrt{2}a$ and $\lambda=\sqrt{3}a$, respectively, instead of \textit{spheres}. An activity $z_k$ is associated with each $k$NN particle (with $k=0,1,2$) in our grand-canonical study of the model. It is worthy recalling that while the pure 0NN model can be exactly solved and it does not present any phase transition, the pure 1NN and 2NN models on the cubic lattice are known to undergo a continuous \cite{Gaunt,Yamagata,HB,Panagiotopoulos} and a discontinuous \cite{Panagiotopoulos} phase transition, respectively, from disordered fluid to ordered solid phases. 

\begin{figure*}[t]
\includegraphics[width=14.0cm]{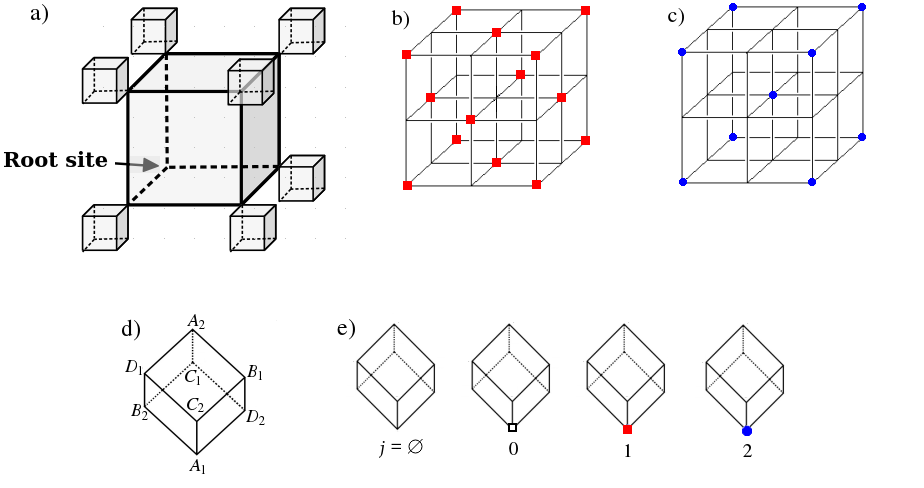}
\caption{(a) Illustration of part of a Husimi lattice built with cubes. The ground states of the solid $S1$ and $S2$ phases, in the cubic lattice, are displayed in (b) and (c), respectively. (d) Definition of the sublattice structure in a elementary cube of the Husimi lattice. (e) Representation of the possible states ($j$) for the root sites. They can be empty ($j=\varnothing$) or occupied by a 0NN ($j=0$, open square), an 1NN ($j=1$, full square) or a 2NN ($j=2$, circle) particle.}
\label{fig1}
\end{figure*}

The translational symmetry breaking of these solid phases display sublattice ordering, as illustrated in Figs. \ref{fig1}(b) and \ref{fig1}(c). In the ground state (the full occupancy limit) the solid phase for the pure 1NN system (the $S1$ phase) is characterized by particles occupying one of two sublattices: either sublattices with index $1$ or $2$ in Fig. \ref{fig1}(d). On the other hand, the solid related to the pure 2NN system (the $S2$ phase) is featured by particles occupying one among four sublattices [$A_i$, $B_i$, $C_i$ or $D_i$, with $i=1,2$, in Fig. \ref{fig1}(d)], so that its ground state is four-fold degenerated. Thereby, to correctly capture the symmetries of both solid phases in the 0NN-1NN-2NN mixture, we have to deal with eight sublattices, as defined in Fig.\ref{fig1}(d).

Following our recent studies on the corresponding binary mixtures \cite{NT0NN2NN,NT0NN1NN-1NN2NN}, here we will investigate the ternary 0NN-1NN-2NN case by defining the model on a Husimi lattice built with cubes [see Fig. \ref{fig1}(a)]. It is important to remark that the Husimi lattice consists of an infinite Cayley tree where each vertice is replaced by a polygon or a polyhedron. So, since the dimension of such lattice is infinity, the critical transition behavior is lead by mean-field exponents. Despite this, solutions on the Husimi lattice carry some degree of correlation of the system on the relevant lattice (the cubic lattice here), usually giving better results than other mean-field methods~\cite{gujrati}. In fact, in some lattice gas systems, the Husimi solution even presents some quantitative agreement with simulation results on regular lattices \cite{Buzano,Tiago10}.

To solve the ternary mixture on the Husimi lattice, we define a root site on an elementary cube [as shown in Fig. \ref{fig1}(a)] and partial partition functions (ppf's) according to its state. For each sublattice, the root site (and any other lattice site as well) can be empty ($j=\varnothing$), occupied by a 0NN ($j=0$), by an 1NN ($j=1$) or by a 2NN particle ($j=2$), totaling $4$ states [see Fig. \ref{fig1}(e)]. Hence, since we have $8$ sublattices, there are $32$ ppf's for this system. One cube defines the $0$-generation of the hierarchical lattice; so, by attaching the root sites of seven cubes to the vertices of such cube, with exception of its root site, one obtains a subtree with $1$-generation. Then, by attaching seven of such subtrees to the vertices of a new cube with exception of its root site, a 2-generation subtree is built. By repeating this process, we can create a $(M+1)$-generation subtree from seven ones with $M$ generations. Then, by summing over all possibilities of creating such subtree, by appropriately taking into account the particle exclusions, with its root site kept fixed in the state $s$ and sublattice $g$, one obtains a recursion relation for the ppf $g'_s$ as function of all ppf's in the previous generation $g_j$, with $g=a_1,a_2,...,d_1,d_2$ and $s,j=\varnothing,0,1,2$. Some details on such recursion relations (RRs) are presented in the appendix. 

Since we are interested in infinite subtrees, and the RRs usually diverges in this thermodynamic limit, we will work with ratios of them, defined as
\begin{equation}
 G_{j}=\frac{g_{j}}{g_{\varnothing}},
  \label{ratio}
\end{equation}
where $G_j=A_{1,j},A_{2,j},...,D_{1,j},D_{2,j}$ and $j=0,1,2$. Thereby, from the $32$ RRs for the ppf's one obtains $24$ RRs for the ratios. The thermodynamic phases of the system are given by the stable and positive fixed points of these RRs. To analyze the stability of a given phase, we calculate the Jacobian matrix and its leading eigenvalue, $\Lambda$, at the associated fixed point. The phase is stable in the regions of the parameter space $(z_0,z_1,z_2)$ where $\Lambda < 1$, while $\Lambda = 1$ defines its spinodal.

Similarly to the ppf's, we can obtain the partition function, $Y$, of the model on the tree by summing over all possibilities of attaching the root sites of eight subtrees to a central cube. It can be written in a compact form in terms of the ppf's, e.g., as
\begin{eqnarray}
 Y&=&a_{1,\varnothing}a'_{1,\varnothing} + z_0 a_{1,0} a'_{1,0} + z_1 a_{1,1}a'_{1,1} + z_2 a_{1,2}a'_{1,2}\\ \nonumber
  &=&  a_{1,\varnothing} b_{1,\varnothing} c_{1,\varnothing} d_{1,\varnothing} a_{2,\varnothing} b_{2,\varnothing} c_{2,\varnothing} d_{2,\varnothing} y,
\end{eqnarray}
where $y$ is a function of the ratios (Eq. \ref{ratio}) and the activities $z_0$, $z_1$ and $z_2$. Then, the density of $k$NN particles in the sublattice $A_1$ at the central cube is given by
\begin{equation}
 \rho^{(A_1)}_k=\frac{A_{1,k}}{8Y}\frac{\partial Y}{\partial A_{1,k}},
\end{equation}
with $k=0,1,2$. A similar equation holds for the other sublattices, by replacing $A_1$ with $A_2,B_1,\ldots$, or $D_2$. Once we have the density of a $k$NN particle in all sublattices, its total density is $\rho_k= \rho^{(A_1)}_k + \rho^{(A_2)}_k + ... + \rho^{(D_1)}_k + \rho^{(D_2)}_k$. From the partition function we can also calculate the bulk free energy per site. Following the ansatz proposed by Gujrati~\cite{gujrati}, and discussed in detail for a Husimi lattice build with cubes in \cite{NT0NN2NN}, for the ternary mixture one has
\begin{equation}
 \phi_b =-\frac{1}{8} \ln \left[ \frac{A_{1,\varnothing} B_{1,\varnothing} C_{1,\varnothing} D_{1,\varnothing} A_{2,\varnothing} B_{2,\varnothing} C_{2,\varnothing} D_{2,\varnothing}}{y^{6}} \right],
\end{equation}
where 
\begin{equation}
 A_{1,\varnothing}\equiv\frac{a'_{1,\varnothing}}{b_{1,\varnothing} c_{1,\varnothing} d_{1,\varnothing} a_{2,\varnothing} b_{2,\varnothing} c_{2,\varnothing} d_{2,\varnothing}},
\end{equation}
and the other ratios can be obtained by the sublattice permutation scheme described in the appendix. The bulk free energy is handy to determine where a first-order transition takes place. In a region where two or more phases have $\Lambda<1$, the equality of their bulk free energies determines the point, line or surface of coexistence among these phases.

\section{Thermodynamic behavior of the model}
\label{therbe}

\subsection{Summary of results for the binary mixtures}

Once the binary mixtures 0NN-1NN ($z_2=0$), 0NN-2NN ($z_1=0$) and 1NN-2NN ($z_0=0$) are the boundary planes of the $(z_0,z_1,z_2$) space for the ternary case, it is interesting to start the presentation of results by discussing in detail the thermodynamic behavior of such planes \cite{NT0NN2NN,NT0NN1NN-1NN2NN}. 

\begin{figure}[!t]
\includegraphics[width=9.0cm]{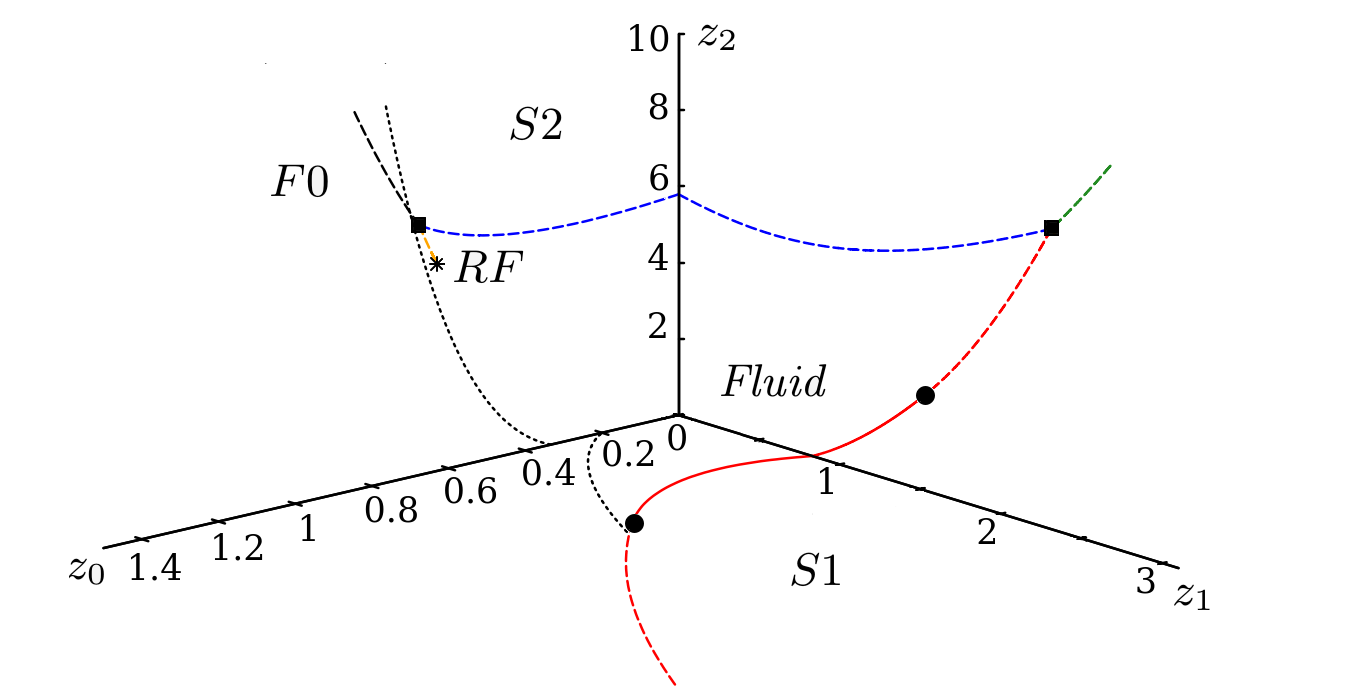}
\caption{Phase diagrams for the binary mixtures in the $(z_0,z_1,z_2)$ space. The solid and dashed lines are continuous and discontinuous transition lines between the indicated phases they are separating. The squares are triple points, the circles are tricritical points and the star is a critical point. The dotted lines are the LMDs.}
\label{fig2}
\end{figure}

For the 0NN-1NN mixture --- the plane $(z_0,z_1,0)$ --- we have found two thermodynamic stable phases \cite{NT0NN1NN-1NN2NN}: a disordered fluid ($F$) phase, where the recursion relations (RRs) for the ratios assume a homogeneous solution $A_{1,j}=B_{1,j}= \ldots = C_{2,j}=D_{2,j}$ with $j=0,1,2$, and the solid $S1$ phase, associated with the ordering of the 1NN particles, where the RRs take the form $A_{i,j}=B_{i,j}=C_{i,j}=D_{i,j}=r_{i,j}$, with $i=1,2$ and $j=0,1,2$, and $r_{1,j}>r_{2,j}$ for $j=0,1$, while $r_{1,2}<r_{2,2}$, when the sublattices with index $1$ are the ones more occupied. Namely, in this phase one has $\rho_j^{(A_1)}=\cdots=\rho_j^{(D_1)}>\rho_j^{(A_2)}=\ldots=\rho_j^{(D_2)}$, for $j=0,1$ and the opposite for $j=2$. [Note that these fixed points and densities are for these phases in the ternary case; and obviously $\rho_2^{(X)}=0$, $\forall$ sublattice $X$, in the 0NN-1NN mixture.] For small $z_0$ there is a continuous $F$-$S1$ transition, while for large $z_0$ such transition becomes discontinuous. The critical and the coexistence $F$-$S1$ lines meet at a tricritical point, located at $(z_{0},z_{1},z_{2})_{TC}=(0.5958,1.1277,0)$ [see Fig. \ref{fig2}]. Interestingly, this mixture presents a thermodynamic anomaly, characterized by minima in isobaric curves of the total density of particles as function of one activity ($z_0$ or $z_1$). A line of minimum density (LMD) exists within the fluid phase, starting at $z_0\approx 0.20$ for $z_1\rightarrow 0$ and ending close to the tricritical point in a region where the $F$ phase is metastable (see Fig. \ref{fig2}).

A similar anomaly has also been observed in the 0NN-2NN mixture --- the plane $(z_0,0,z_2)$. In this case, the LMD starts at $z_0\approx 0.333$ for $z_2\rightarrow 0$ and, as in the 0NN-1NN case, it ends inside the region where the fluid phase is metastable~\cite{NT0NN2NN} (see Fig. \ref{fig2}). Actually, beyond the regular fluid ($RF$) phase (whose RRs have the symmetry just mentioned for the $F$ phase), the 0NN-2NN mixture displays another stable disordered fluid phase characterized by a dominance of 0NN particles, reason for which it was baptized as the $F0$ phase in Ref. \cite{NT0NN2NN}. The RRs are also featured by $A_{1,j}=B_{1,j}= \ldots = C_{2,j}=D_{2,j} = r_j$ in the $F0$ case, but with $r_1 \approx 1$, $r_2 \approx 0$ and $r_3 \approx 0$, while in $RF$ case the values of $r_j$ strongly depend on the activities. These very same symmetries apply for the particle densities: while in $RF$ phase $\rho_0$, $\rho_1$ and $\rho_2$ considerably vary with $z_0$, $z_1$ and $z_2$, in the stable region of the $F0$ phase one has $\rho_0 \gg \rho_1$ and $\rho_0 \gg \rho_2$. Beyond these two fluid phases, the solid $S2$ phase is also present in the phase diagram, which is associated with the ordering of the 2NN particles and characterized by a fixed point with $A_{1,j}=A_{2,j}>B_{1,j}=C_{1,j}=\ldots=C_{2,j}=D_{2,j}$, for $j=0,1,2$, when the sublattices $A_1$ and $A_2$ are the more populated ones. This yields densities $\rho_j^{(A_1)}=\rho_j^{(A_2)} > \rho_j^{(B_1)}=\ldots=\rho_j^{(D_2)}$, for $j=0,1,2$. A stable fluid-fluid demixing is observed in this system, whose discontinuous $RF$-$F0$ transition line ends at a critical point (CP), located at $(z_{0},z_{1},z_{2})_{CP}=(0.6297,0,5.4243)$, so that for $z_2<z_{2,CP}$ the $RF$ and $F0$ phases cannot be distinguished. Hereafter, whenever this happens we will refer to these phases simply as the fluid ($F$) phase. The $S2$ phase is the most stable one for large $z_2$ and it is separated from the two fluid phases by discontinuous transition lines. The three coexistence lines $RF$-$F0$, $RF$-$S2$ and $F0$-$S2$ meet at a triple point (TP) located at $(z_{0},z_{1},z_{2})_{TP}=(0.6774,0,6.5671)$ [See Fig. \ref{fig2}].

Finally, in the 1NN-2NN mixture --- the plane $(0,z_1,z_2)$ ---, we have found the fluid $F$ and the two solid $S1$ and $S2$ phases \cite{NT0NN1NN-1NN2NN}. Similarly to the 0NN-1NN mixture, continuous and discontinuous $F$-$S1$ transition lines are observed, which meet at a tricritical point, located at $(z_{0},z_{1},z_{2})_{TC}=(0,1.5273,2.5016)$. The solid $S1$ and $S2$ phases are observed in the regions of large $z_1$ and $z_2$, respectively, and are separated by a discontinuous transition line. Another coexistence line exists between the $F$-$S2$ phases, which meet the $F$-$S1$ and $S1$-$S2$ ones in a triple point, located at $(z_{0},z_{1},z_{2})_{TP}=(0,2.3102,7.8746)$ [See Fig. \ref{fig2}]. In contrast with the previous binary mixtures, no density anomaly was found in the 1NN-2NN case.

The summary of the behavior for the binary mixtures, displayed in Fig. \ref{fig2}, indicates that the 3D phase diagram for the ternary case is quite complex and interesting. For instance, the presence of the two fluid-$S2$ discontinuous transition lines in the planes $(0,z_1,z_2)$ and $(z_0,0,z_2)$ strongly suggests the existence of a coexistence fluid-$S2$ surface in the 3D parameter space. Similarly, the $F$-$S1$ transitions in the planes $(z_0,z_1,0)$ and $(0,z_1,z_2)$ strongly indicates that a critical and a coexistence $F$-$S1$ surface exist in the full phase diagram, both meeting at a line of tricritical points (a TC line). Besides these features, everything else is less clear. For example, it is unclear from the binary mixtures what can be expected for the LMD lines found in the planes $(z_0,z_1,0)$ and $(z_0,0,z_2)$, since the values of $z_0$ where they start in each plane do not coincide. So, this could indicate either a complex scenario with two (or more) surfaces of minimum density, or that one or both of such surfaces do not exist. In the same token, it is not possible to known at this point whether the fluid-fluid demixing is a particular feature of the case $z_0=0$, or if it extends for $z_0>0$, giving rise to a $RF$-$F0$ coexistence surface, as well as to a line of triple points (where the phases $RF$-$F0$-$S2$ coexist) and a critical line in the 3D phase diagram. In what follows, these points will be addressed in great detail.

\subsection{The Fluid-Fluid transition}

From all thermodynamic properties observed in the binary mixtures discussed above, the fluid-fluid demixing transition in the 0NN-2NN ($z_1=0$) system is the most surprising one, in face it is absence in other $k$NN mixtures studied so far. For this reason, we will start the building up of the 3D phase diagram for the ternary mixture by analyzing the $RF$-$F0$ transition.

\begin{figure}[!t]
\includegraphics[width=8.0cm]{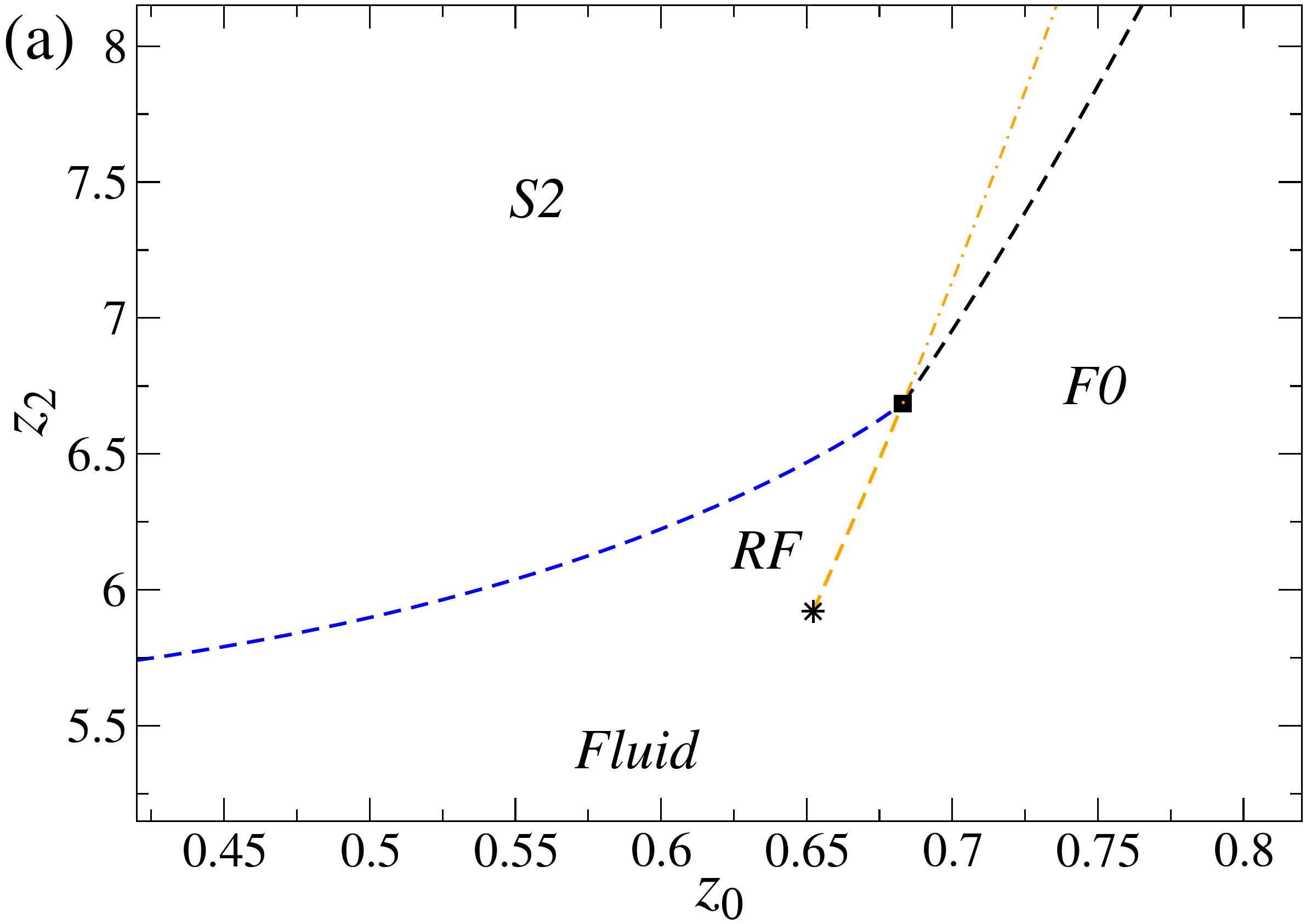}
\includegraphics[width=8.0cm]{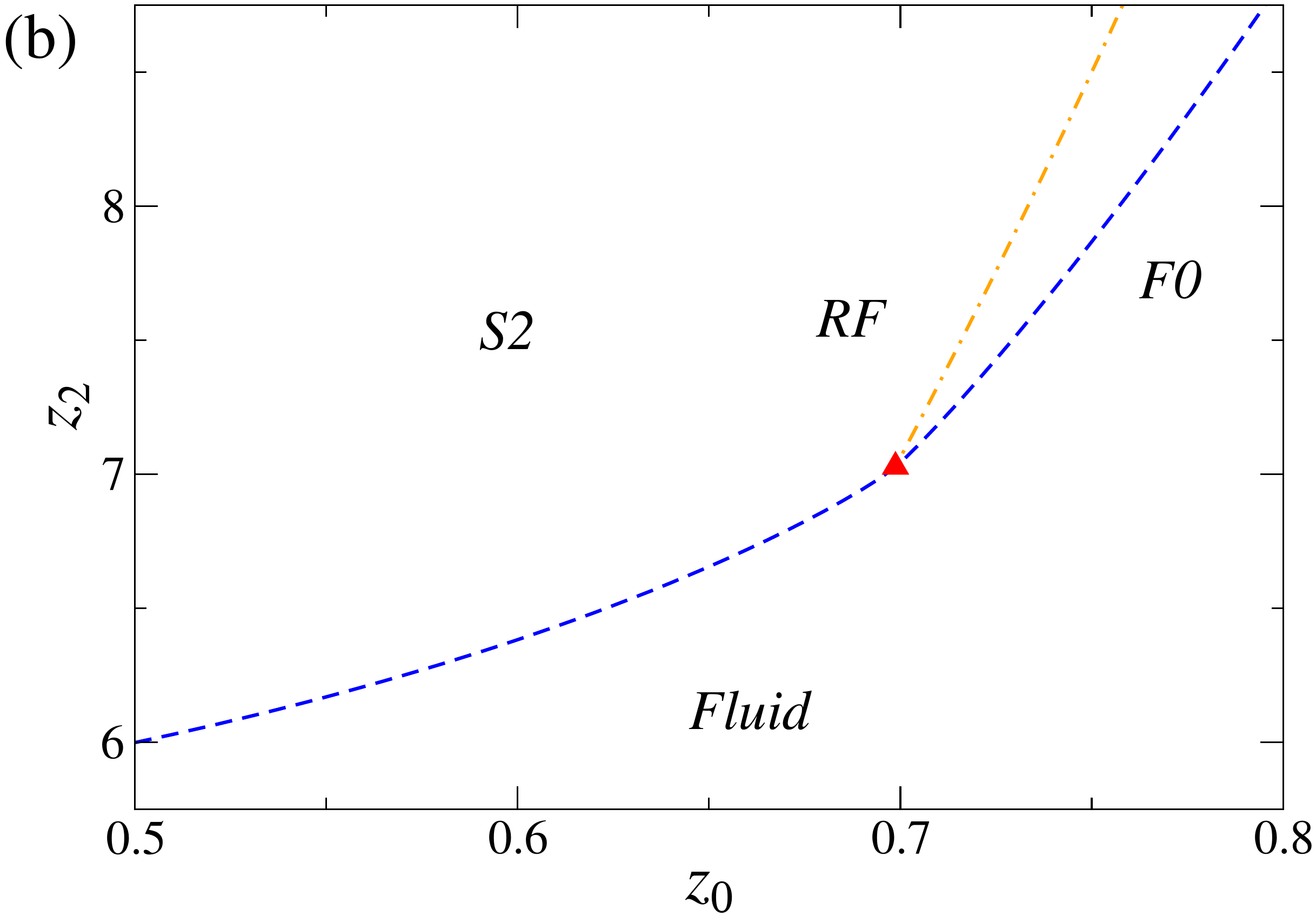}
\includegraphics[width=8.0cm]{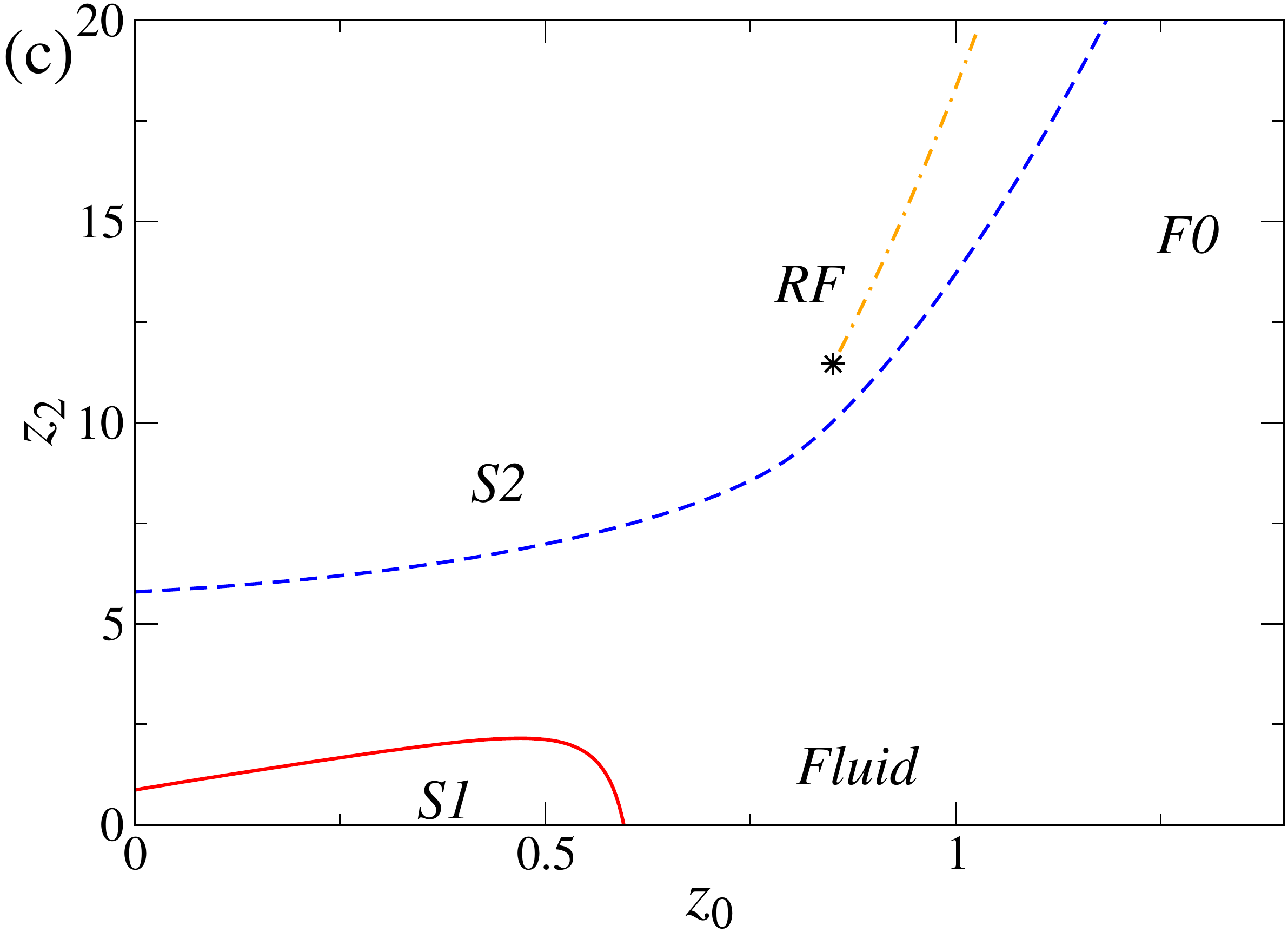}
\caption{Phase diagrams for fixed (a) $z_1=0.10$, (b) $z_1=0.3177$ and (c) $z_1=1.10$. In all panels, the dashed lines represent \textit{stable} coexistence lines between the phases they are separating, while the dashed-dotted lines are the \textit{metastable} $RF$-$F0$ coexistence lines. The solid line in (c) is the $F$-$S1$ critical line. The square in (a) is the $RF$-$F0$-$S2$ TP, while the stars are the $RF$-$F0$ CP and the triangle in (b) represents the special point where the TP line ends.}
\label{fig3}
\end{figure}

In order to do this, we investigate the phase behavior of slices of the 3D space for fixed values of $z_1$. Some relevant examples of such slices are shown in Fig. \ref{fig3}. For small $z_1$, specifically for $z_1<0.3177$ [see Fig. \ref{fig3}(a) for the case $z_1=0.1$], the same thermodynamic behavior of the case $z_1=0$ is found, with the phases $RF$, $F0$ and $S2$ pairwise separated by discontinuous transition lines which meet at a triple point, while the fluid-fluid demixing transition ends in a critical point. This demonstrates that in the 3D phase diagram there exist coexistence surfaces separating the $RF$-$F0$, $RF$-$S2$ and $F0$-$S2$ phases, all of them meeting at a line of triple points (the $RF$-$F0$-$S2$ TP line). Moreover, there is a line of critical points (a CP line), where the $RF$-$F0$ coexistence surface ends. Substantially, this confirms that the fluid-fluid demixing transition is not restricted to the $z_1=0$ case, being a feature of the ternary mixture.

Quantitatively, we observe that while the $(z_0,z_2)$ coordinates of the TP line mildly change with $z_1$, the $z_2$ one for the CP line present a strong variation with $z_1$, leading the stable region of the $RF$-$F0$ coexistence to decrease as $z_1$ increases. For instance, the $RF$-$F0$ coexistence line occurs in an interval $\Delta z_2\approx1.14$ for $z_1=0$, but only $\Delta z_2 \approx 0.76$ for $z_1=0.1$. In fact, by increasing $z_1$ one observes that the CP line becomes closer to the TP line, and at the special point $(z_0,z_1,z_2)^*=(0.6987,0.3177,7.0280)$ both lines meet, so that the \textit{stable} fluid-fluid demixing transition disappears at this point. This gives rise to the phase diagram displayed in Fig. \ref{fig3}(b), for $z_1=z_1^*$, with the (now completely metastable) $RF$-$F0$ coexistence line ending at the special point, so that the $RF$-$S2$ coexistence surface for $z_1<z_1^*$ now becomes a $F$-$S2$ coexistence, which meets the $F0$-$S2$ coexistence line at the special point. 

\begin{figure}[!t]
\includegraphics[width=8.5cm]{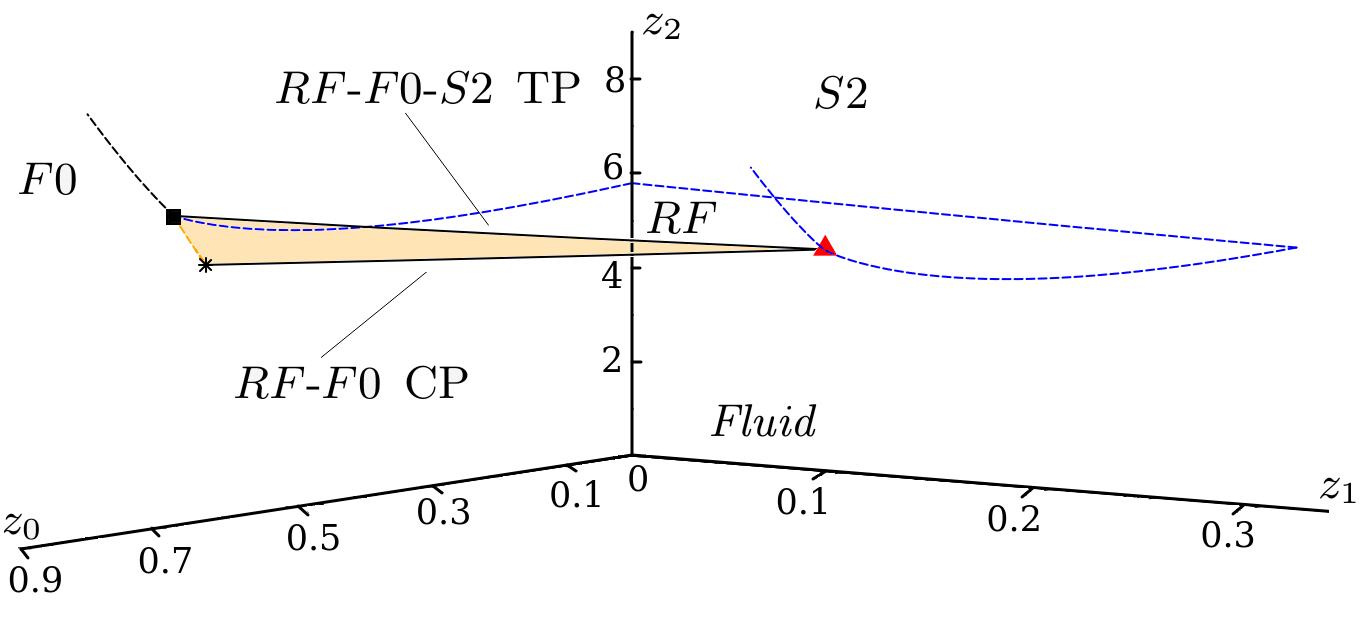}
\caption{Part of the 3D phase diagram for the ternary mixture in the activity space, highlighting the region around which the fluid-fluid demixing surface [the shaded (orange) one in the plot] takes place. All the surfaces presented here (separating the phases $RF$- or $F$-$S2$, $F0$-$S2$ and $RF$-$F0$) are coexistence surfaces. The triangle is the special point, where the CP line meets the $RF$-$F0$-$S2$ TP line. The star and the square represent the CP and the TP point, respectively, in the $(z_0,0,z_2)$ plane.}
\label{fig4}
\end{figure}

For not so large values of $z_1>z_1^*$, one still finds the $RF$-$F0$ coexistence ending at the CP line, as shows Fig. \ref{fig3}(c) for $z_1=1.1$, but now this line is also metastable, occurring inside the region where the $S2$ phase is the most stable one. Therefore, the more complex scenario observed for small $z_1$ gives place to a simple $F$-$S2$ discontinuous transition. Namely, for $z_1>z_1^*$ the fluid-fluid demixing transition becomes preempted by the fluid-solid transition. Hence, at the special point the CP line becomes metastable and the $RF$-$F0$-$S2$ TP line ends. A summary of these behaviors in the 3D space is presented in Fig. \ref{fig4}.

\begin{figure}[!t]
\includegraphics[width=8.0cm]{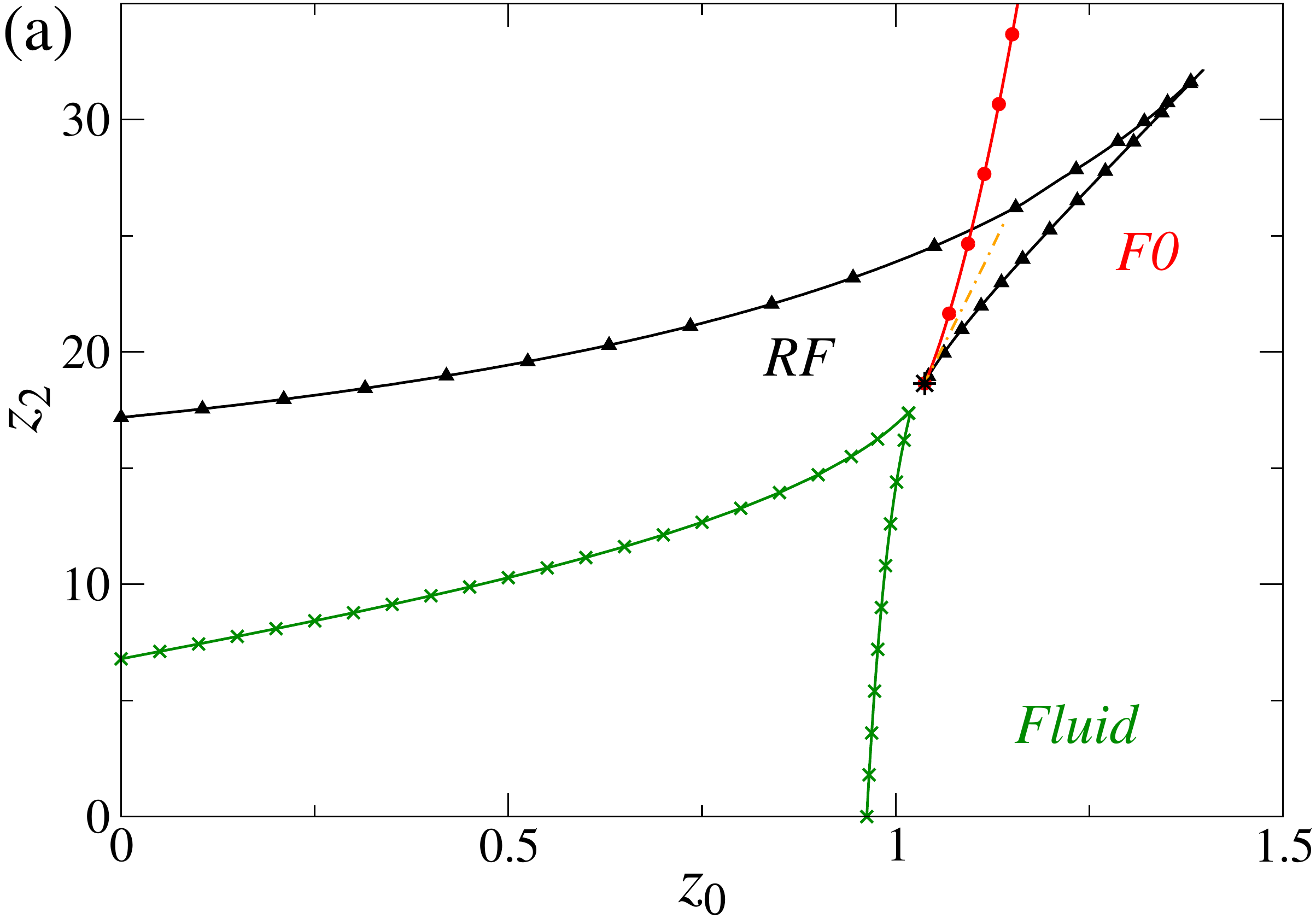}
\includegraphics[width=8.0cm]{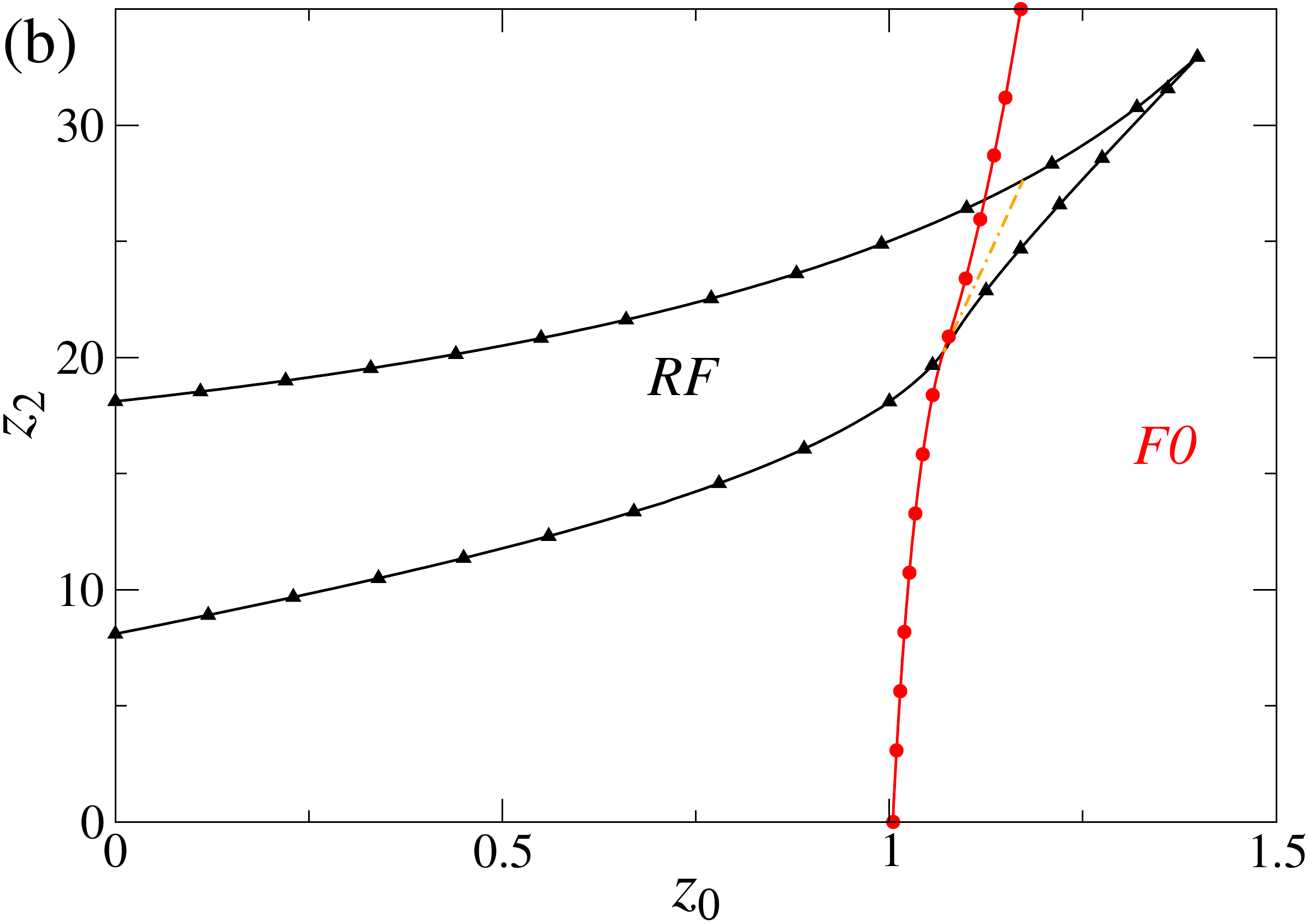}
\includegraphics[width=8.0cm]{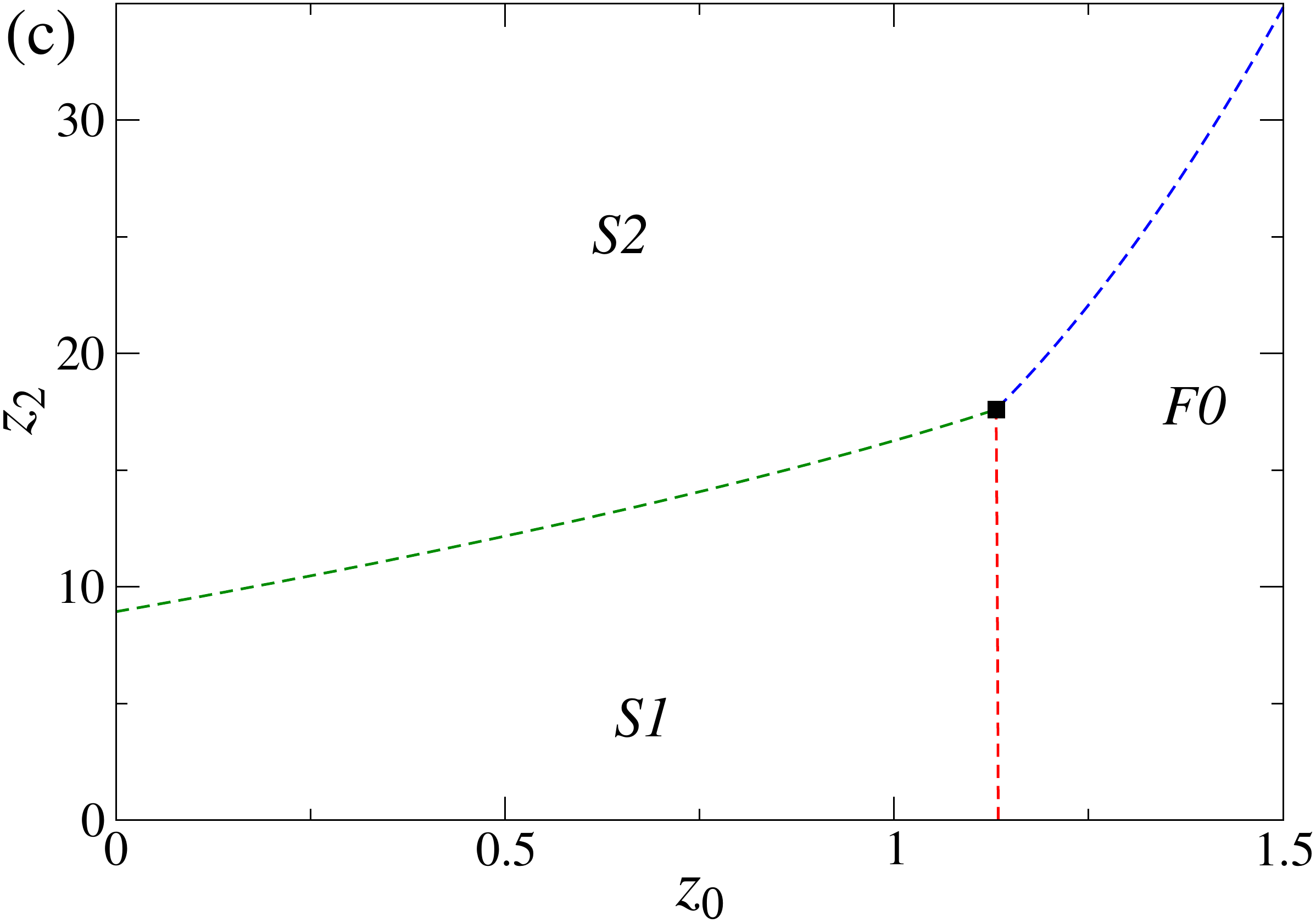}
\caption{Spinodals of the disordered fluid phases (solid lines) and the metastable $RF$-$F0$ coexistence lines (dash-dotted lines) for fixed (a) $z_1=2.3$ and (b) $z_1=2.5$. The lines with triangles, circles and crosses are the $RF$, $F0$ and fluid (where $RF$ and $F0$ phases cannot be distinguished) spinodals, respectively. The star in (a) is the metastable CP. For comparison, the phase diagram for $z_1=2.5$ is presented in (c), where the dashed lines represent coexistence lines between the indicated phases they are separating and the square is the TP. The phase diagram for $z_1=2.3$ is quite similar to this one.}
\label{fig5}
\end{figure}

Although in general we will not be interested in discussing metastable transitions here, it is interesting to take a close look on this for the fluid-fluid demixing, to verify how further the $RF$-$F0$ coexistence surface goes as $z_1$ increases. First, we notice that, regardless the value of $z_1$, the metastable part of this surface is always limited from above, i.e., it ends when it meets the spinodal of the $RF$ phase, as shown in Figs. \ref{fig5}(a) and \ref{fig5}(b). Second, for large $z_1$ the $S1$ phase also appears in the phase diagram in the region of small $z_0$ and $z_2$, as seen in Fig. \ref{fig3}(c), and the fluid phase let to be stable in such region. This is clearly seen by comparing the phase diagram in Fig. \ref{fig5}(c) with the spinodals of the fluid phases in Fig. \ref{fig5}(b). Curiously, by increasing $z_1$ the spinodal of the fluid phase develops a cusp, which approximates of the (metastable) CP line as $z_1$ increases, see Fig. \ref{fig5}(a) for $z_1=2.3$. Such approximation occurs until $z_1 \approx 2.3731$, after which the cusp meets the $RF$ and $F0$ spinodals (in the point where there was the CP line), changing completely the scenario of such spinodals. In fact, as demonstrated in Fig. \ref{fig5}(b), for $z_1=2.5$, the stability region of $RF$ phase becomes now limited to a closed domain of the ($z_0,z_2$) phase diagram, and the (metastable) $RF$-$F0$ coexistence line ends at the point where the spinodal of the $F0$ phase crosses the one for the $RF$ phase. Therefore, the CP line ends at $z_1 \approx 2.3731$, while the fluid-fluid demixing still exists for larger values of $z_1$. However, by increasing $z_1$, one observes that the domain of the $RF$ stability shrinks, yielding a decreasing in the in $RF$-$F0$ coexistence line, which finally disappears at $z_1=5.7157$.

\subsection{The phase diagram for the ternary mixture}

From the results in the previous subsection, one knows that for $z_1<z_1^*$ the 3D phase diagram presents three coexistence surfaces ($RF$-$F0$, $RF$-$S2$ and $F0$-$S2$), a $RF$-$F0$-$S2$ TP line and a CP line, which meet (and ends or becomes metastable) at the special point, as depicted in Fig. \ref{fig4}. Moreover, for $z_1>z_1^*$, but not so large, this behavior gives place to a simple $F$-$S2$ coexistence surface.

\begin{figure}[!t]
\includegraphics[width=8.0cm]{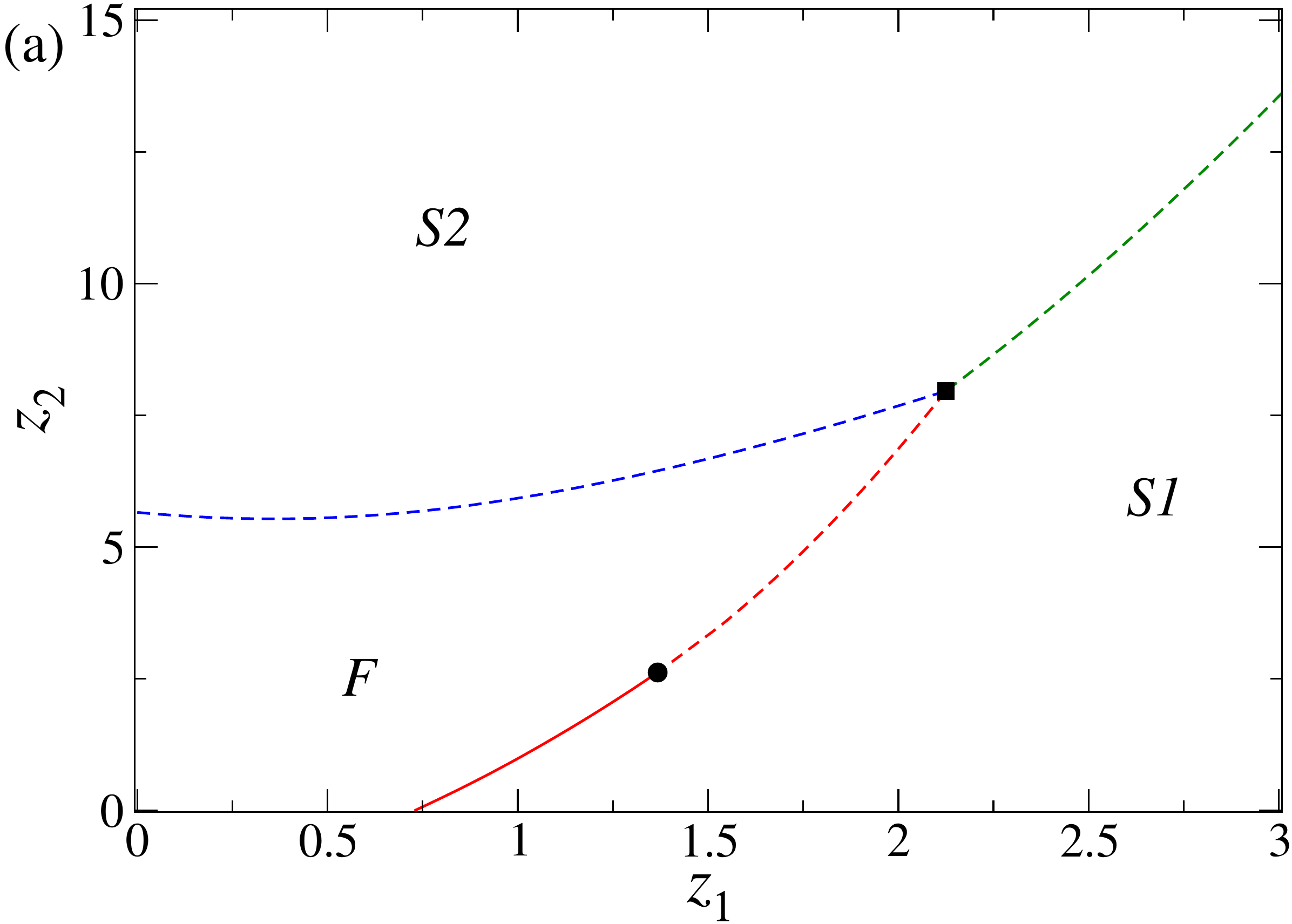}
\includegraphics[width=8.0cm]{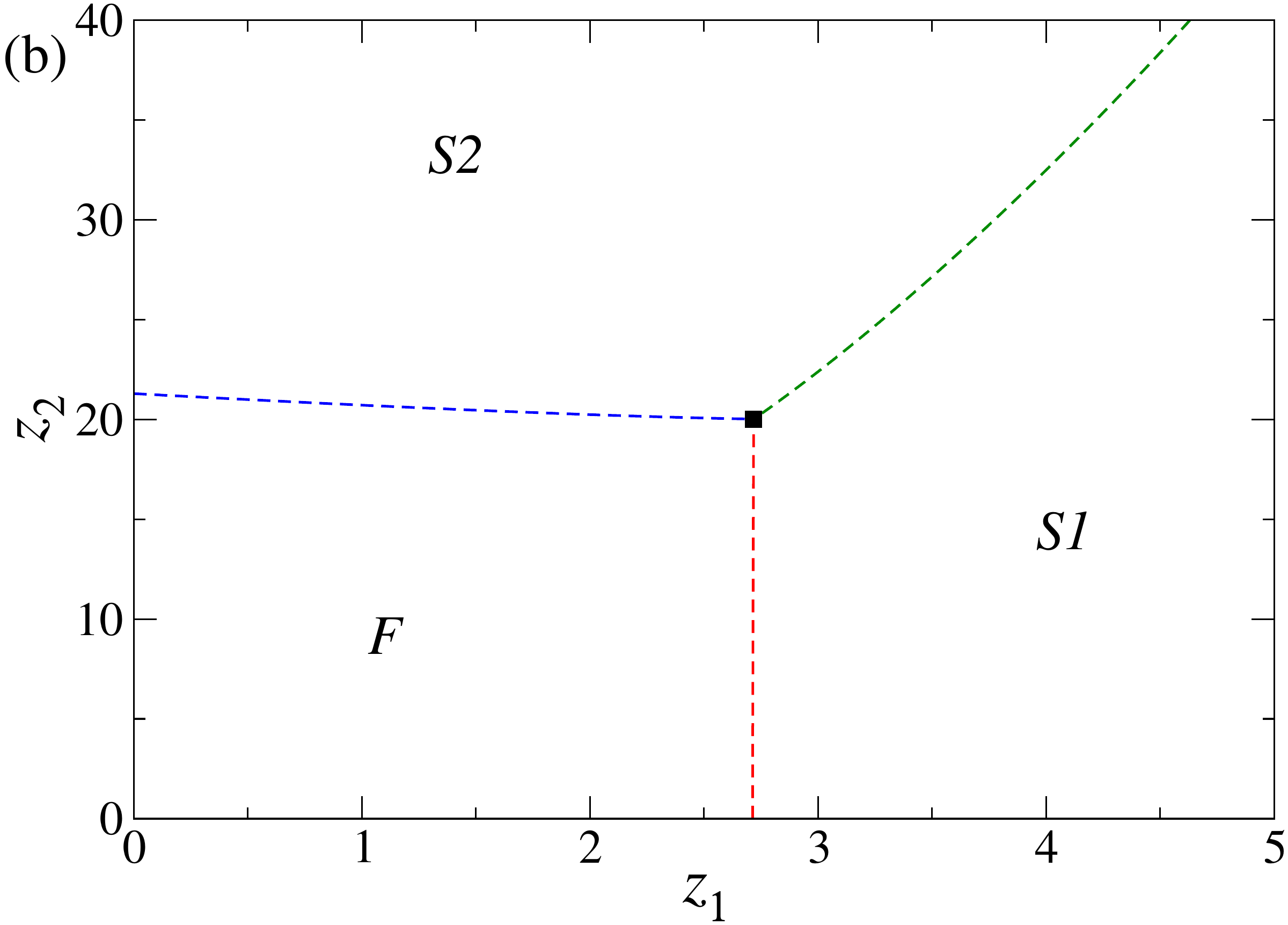}
\caption{Phase diagrams for fixed (a) $z_0=0.20$ and (b) $z_0=1.20$. The solid and dashed lines represent critical and coexistence lines, respectively, between the indicated phases they are separating. The black squares and the circle denote the TP and the TC point, respectively.}
\label{fig6}
\end{figure}

\begin{figure}[!t]
\includegraphics[width=8.5cm]{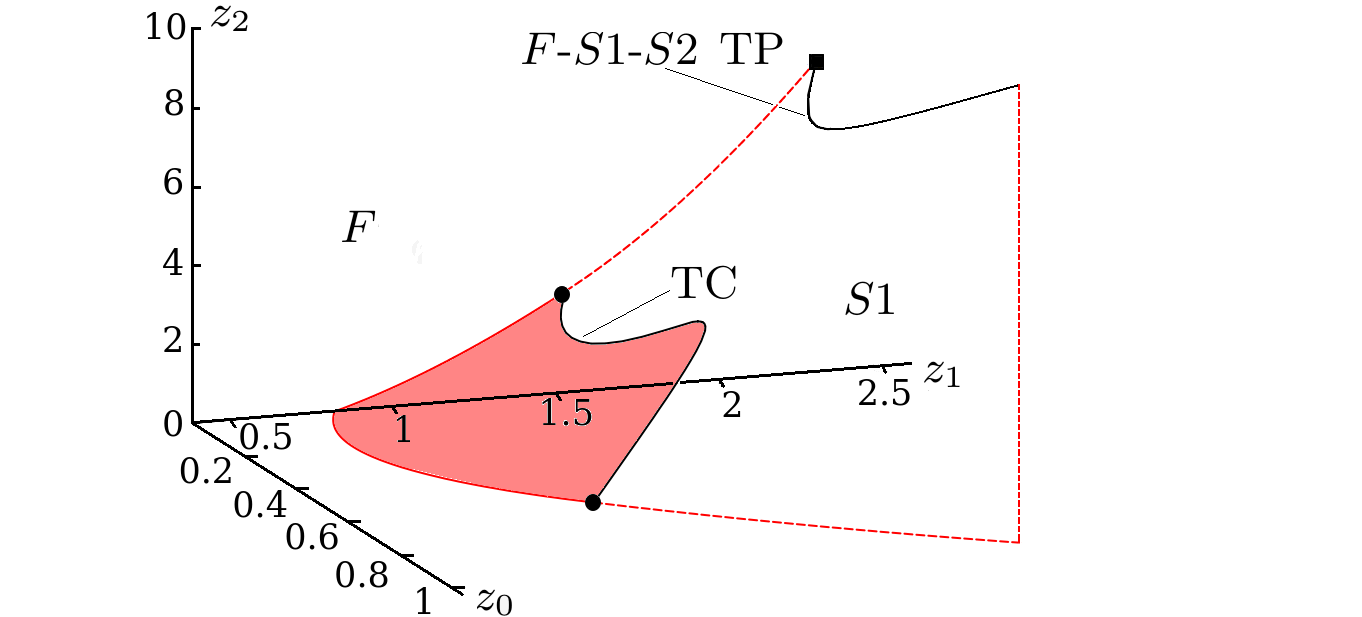}
\caption{Part of the 3D phase diagram for the ternary mixture in the activity space, highlighting the region where the critical $F$-$S1$ surface [the shaded (red) one in the plot] takes place. The the dashed lines indicate the $F$-$S1$ coexistence surface, which meets the critical surface at the TC line and is limited from above by the $F$-$S1$-$S2$ TP line. The circles and the square mark the location of the TC line and TP line, respectively, in the boundary planes.}
\label{fig7}
\end{figure}

As shows Fig. \ref{fig3}(c), a critical $F$-$S1$ transition line is found in slices of the phase diagram for $z_1$ fixed (and large enough), giving rise to a critical $F$-$S1$ surface in the 3D space, as expected from the behavior of the binary mixtures. Such surface starts at $z_1=0.7284$ --- which turns out to be the point of minimum (in relation to $z_1$) of the $F$-$S1$ critical line in the 0NN-1NN mixture [i.e., in the plane ($z_0,z_1,0$)] \cite{NT0NN1NN-1NN2NN} --- and exists in a limited region of the parameter space. Namely, it gives place to a discontinuous $F$-$S1$ transition surface for large $z_1$, as confirmed in Fig. \ref{fig5}(c). These features are better seen in slices of fixed and small $z_0$, as the one in Fig. \ref{fig6}(a) for $z_0=0.2$, where one finds a phase diagram similar to the one for $z_0=0$, with a continuous and a discontinuous $F$-$S1$ transition line meeting at a TC point. This confirms not only the existence of a critical and a coexistence $F$-$S1$ surface, but also the presence of a TC line where they meet in the 3D space. By analyzing several slices for fixed $z_0$ or $z_2$, we have accurately determined the TC line, which connects the two tricritical points found in the planes $(z_0,z_1,0)$ and $(0,z_1,z_2)$. Interestingly, it presents a complex non-monotonic behavior, attaining a maximum point, with respect to $z_1$ and $z_2$, at $(z_{0},z_{1},z_{2})_{TC,max}=(0.6719,1.4216,4.6678)$. This line, as well as the critical and coexistence $F$-$S1$ surfaces are depicted in Fig. \ref{fig7}, which shows this part of the 3D phase diagram in detail.

The $z_0$-slices, as those in Figs. \ref{fig6}(a) and \ref{fig6}(b), also confirm the presence of the discontinuous $F$-$S2$ surface, as well as unveil the existence of a surface of solid-solid ($S1$-$S2$) demixing in the ternary mixture. Moreover, one always observes that the $F$-$S1$, $F$-$S2$ and $S1$-$S2$ coexistence surfaces meet in a line of triple points (the $F$-$S1$-$S2$ TP line), as seen in Figs. \ref{fig5}(c) and \ref{fig6}. This line starts at the triple point of the 1NN-2NN mixture [i.e., in the ($0,z_1,z_2$) plane] and extends to $z_0,z_1,z_2\rightarrow\infty$; its behavior is also shown in Fig. \ref{fig7}. It is important to notice that, since the TC line (and then the critical $F$-$S1$ surface) ends at a finite value of $z_0$, for large values of this activity everything we find in this part of the 3D phase diagram are the $F$-$S1$, $F$-$S2$ and $S1$-$S2$ coexistence surfaces meeting at the TP line [see Fig. \ref{fig6}(b)].

The existence of the $F$-$S1$-$S2$ TP line in the limit of $z_0,z_1,z_2\rightarrow\infty$ can be justified by the existence of a $F0$-$S1$, a $F$-$S2$ and a $S1$-$S2$ coexistence line in the binary mixtures, all of them extending to infinity \cite{NT0NN2NN,NT0NN1NN-1NN2NN}. Since the RRs assume simple forms in such limit, it is straightforward to determine the behavior of the TP line there. For instance, the fixed point for the RRs in the fluid phase has the form $A_{i,0}=B_{i,0}=C_{i,0}=D_{i,0}=1$ and $A_{i,j}=B_{i,j}=C_{i,j}=D_{i,j}=0$ for $i=1,2$ and $j=1,2$. For the $S1$ phase one finds $A_{1,j}=B_{1,j}=C_{1,j}=D_{1,j}=1$ for $j=0,1$, while all others RRs vanish, when the sublattices indexed by $1$ are the more populated ones. In the $S2$ phase, if the sublattices $A_1$ and $A_2$ are the ones more populated, one has $A_{i,j}=1$ and $B_{i,j}=C_{i,j}=D_{i,j}=0$, for $i=1,2$ and $j=0,1,2$. Using these solutions, it is simple to calculate the free energies in this limit, being $\phi^{(F)}=(1+z_0)^8$, $\phi^{(S1)}=(1+z_0+z_1)^4$ and $\phi^{(S2)}=(1+z_0+z_1+z_2)^2$. By making $\phi^{(F)}=\phi^{(S1)}=\phi^{(S2)}$ and solving in terms of $z_0$, one finds
\begin{equation}
 z_1=z_0^2 + z_0
\end{equation}
and
\begin{equation}
 z_2=z_0^4 + 4z_0^3 + 5z_0^2 + 2 z_0.
\end{equation}
Therefore, as $z_0\rightarrow\infty$ this TP line behaves has $z_1\approx z_0^2$ and $z_2\approx z_0^4$. This result can be understood as follows: at full occupancy an 1NN particle effectively occupies the volume of two 0NN ones, while a 2NN particle occupies the volume of four 0NN ones, in agreement with the exponents above. The calculated free energies also allow us to determine the behavior of the $F$-$S1$, $F$-$S2$ and $S1$-$S2$ coexistence surfaces in the limit of $z_0,z_1,z_2\rightarrow\infty$, where one finds $z_1\approx z_0^2$ and $z_2\approx z_0^4$ for the $F$-$S1$ and $F$-$S2$ cases, respectively. Therefore, in slices of the 3D phase diagram for very large and fixed $z_0$, these surfaces shall appear as straight lines at constant $z_1$ and $z_2$, respectively. The $S1$-$S2$ coexistence surface, on the other hand, behaves as $z_2\approx (z_0+z_1)^2$, being a quadratic function of $z_1$ ($z_0$) for fixed $z_0$ ($z_1$). In fact, these results are somewhat confirmed in Fig.\ref{fig6}(b) for $z_0=1.2$, where we see that the $F$-$S1$ and $F$-$S2$ lines occur at almost constant values (specially the former one), while the $S1$-$S2$ line is not so straight, indicating the existence of a nonlinear dependence between $z_2$ and $z_1$. Results [not shown] for slices for much larger values of $z_0$ indeed confirm the correctness of the predicted surfaces.

\begin{figure}[!t]
\includegraphics[width=8.5cm]{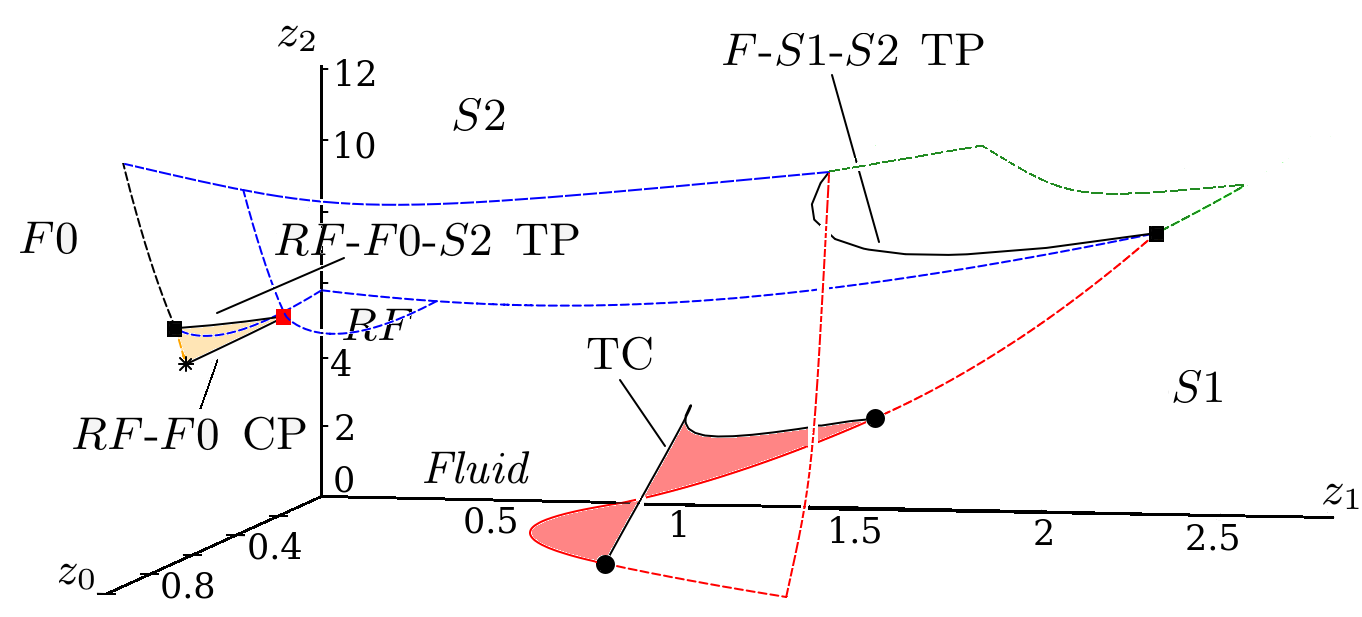}
\caption{Phase diagram for the ternary mixture in the activity space, indicating all phases and transition surfaces and lines, following the same scheme of symbols and colors from the previous figures.}
\label{fig8}
\end{figure}

Figure \ref{fig8} presents the complete 3D phase diagram in the space of activities, summarizing all the features discussed so far. In face of its complexity --- it has a fluid-fluid, a solid-solid and several fluid-solid coexistence surfaces, with three of them ($F$-$S1$, $F$-$S2$, and $S1$-$S2$) extending to $z_0,z_1,z_2\rightarrow \infty$, beyond a critical fluid-solid surface, a CP line, a TC line and two TP lines --- it is a bit hard drawing this phase diagram in an intelligible way. Anyhow, it is important to let clear how the partial diagrams depicted in Figs. \ref{fig4} and \ref{fig7} connects, what are the sizes of the transition surfaces (at least the ones which exist in limited domains) and so on. 

At this point, it is important to stress that we have carefully looked for new phases in the general case, but only the four phases already present in the binary mixtures were found.

\subsection{The surfaces of minimum density}

As discussed in subsection \ref{therbe}A, two lines of minimum density (LMDs) were found in the binary mixtures \textit{inside the fluid phase}, one in the 0NN-1NN ($z_2=0$) and other in the 0NN-2NN ($z_1=0$) case. In the former system, the LMD starts at $z_0 \approx 0.333$ when $z_1\rightarrow 0$, while in the last one it starts at $z_0 \approx 0.20$ when $z_2 \rightarrow 0$, and in both cases they end at the spinodal of the fluid phase, inside regions where the corresponding solid phases are more stable than the fluid \cite{NT0NN1NN-1NN2NN,NT0NN2NN}. The difference between these starting points suggests a complex scenario for the ternary mixture with at least two surfaces of minimum density (SMDs). Before discussing them, however, it is important to remark that such starting points were calculated through a limiting process. For example, for the 0NN-1NN mixture, we located the ($z_0,z_1$) coordinate where the total density --- defined as $\rho_T=\rho_0 + 2\rho_1 + 4 \rho_2$ --- presents a minimum for a given pressure, $P$ \footnote{In our grand-canonical formalism $P=-\phi/a^3$.}, i.e., along an isobaric curve of $\rho_T \times z_0$ (or $z_1$) (see, e.g., Fig. 5 in Ref. \cite{NT0NN2NN}). Then, by varying $P$ we can build up the LMD curve and determine it as close as we want of the $z_0$ axis by making $z_1 \rightarrow 0$. Exactly on the $z_0$-axis, however, one has $\rho_T=z_0/(1+z_0)$, so that no anomaly exists there. 

\begin{figure}[!t]
\includegraphics[width=8.5cm]{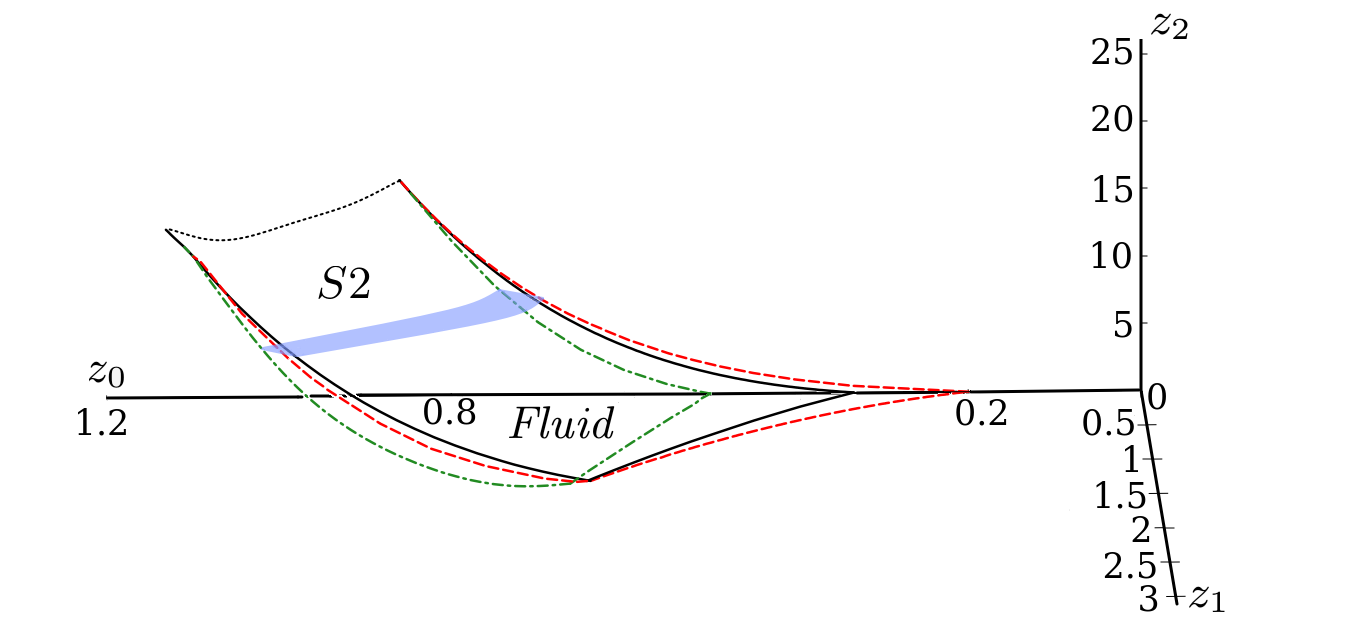}
\caption{Surfaces of minimum density in the activity space. The dash-dotted (green), continuous (black) and dashed (red) lines are the $z_0$-, $z_1$- and $z_2$-SMD, respectively. The doted (black) line is the LMD where all three SMDs seems to end. The shaded (blue) surface is part of the $F$-$S2$ coexistence surface, above which the SMDs become metastable.}
\label{fig9}
\end{figure}

Then, if we take slices of fixed $z_2>0$, one still finds LMDs in the way just described, forming a SMD in the 3D space. Such $z_2$-SMD (since it is obtained for fixed values of $z_2$) starts at the LMD in the plane $(z_0,z_1,0)$ and varies smoothly in the space, ending at the spinodal of the fluid phase, deep inside the region where the fluid is metastable. Although this surface does not exist exactly in the plane $(z_0,0,z_2)$, using the limiting process we can obtain it as close as we want of this plane, so that there exists a LMD for $z_1 \rightarrow 0$ limiting this surface (see Fig. \ref{fig9}). Proceeding in the same way, but now considering slices of fixed $z_1$, a second SMD (the $z_1$-SMD) is obtained, which starts in the LMD on the plane $(z_0,0,z_2)$ and also ends at the spinodal of the fluid phase in the metastable region. In this case, the SMD does not exist exactly in the plane $(z_0,z_1,0)$, but it can be determined in the limit of $z_2 \rightarrow 0$. This $z_1$-SMD is also shown in Fig. \ref{fig9}. The existence of these two SMDs indicates, for sake of completeness, the existence of a third one for fixed $z_0$. As shown in Fig. \ref{fig9}, such $z_0$-SMD indeed exists in the 3D space and, albeit it cannot be observed exactly in the planes $(z_0,z_1,0)$ and $(z_0,0,z_2)$, it can be obtained in the limits $z_2 \rightarrow 0$ and $z_1 \rightarrow 0$, respectively. In such limits, the starting point of the $z_0$-SMD is at $z_0 \approx 0.50$ and, similarly to the other SMDs, it also ends in the spinodal of the fluid phase. As observed in Fig. \ref{fig9}, the three SMDs become quite close for large $z_2$ and they seem to be limited from above by a single line. However, it is quite hard to assure this numerically and it may be the case that they end at different (but very close) lines, or become a single surface before this.

\section{Conclusions}
\label{SecConclusions}

We have determined the thermodynamic behavior of a ternary mixture of hard spheres, defined on the cubic lattice, composed by point-like particles (0NN) and particles which exclude up to their first (1NN) and second (2NN) neighbors. We treat this model by solving it on a Husimi lattice built with cubes, where four phases were found in the phase diagram, being two fluid and two solid ones, separated by several coexistence surfaces and a critical surface, which end or meet in lines of critical, tricritical and triple points. One of the disordered phases is a regular fluid ($RF$), in the sense that the particle densities ($\rho_0$, $\rho_1$ and $\rho_2$) strongly depends on the activities, while in the second disordered fluid ($F0$) phase one always has $\rho_0 \gg \rho_1$ and $\rho_0 \gg \rho_2$. The solid $S1$ ($S2$) phase is featured by a sublattice ordering of 1NN (2NN) particles. We notice that columnar and smectic phases, observed for hard cubes on the cubic lattice \cite{Lafuente,Rajesh}, are absent in our solution. Although this can be due to the hierarchical structure of the Husimi lattice, we remark that these phases have never been found in previous studies of the pure 1NN and 2NN models on the cubic lattice \cite{Gaunt,Yamagata,HB,Panagiotopoulos}, strongly suggesting that they should indeed not appear in their mixture.   

A fluid-fluid ($RF$-$F0$) demixing surface is observed in the system, which is limited from below [in the ($z_0,z_1,z_2$) space] by a line of critical points (CP), whose coordinate $z_{2,c}$ increases fast with $z_{1,c}$, while the CP line and the whole $RF$-$F0$ coexistence surface exist in a narrow range of $z_0$. This shows that, by increasing $\rho_1$, a large $\rho_2$ is need to yield the demixing, confirming that the fluid-fluid transition is not a feature restricted to the 0NN-2NN mixture, but the presence of the 1NN particles turns its appearance more difficult. In fact, although the demixing surface extends for a large portion of the phase diagram, it is stable only in a small area, when compared with the other transition surfaces. Such stable part is limited from above by the $RF$-$F0$-$S2$ triple point line, which ends when it meets the CP line at a special point [located at $(z_0,z_1,z_2)^*=(0.6987,0.3177,7.0280)$]. Therefore, the demixing surface becomes completely metastable (i.e., preempted by the $F$-$S2$ transition) already for $z_1>z_1^*$, corresponding to a very small density of 1NN particles: $\rho_1>\rho_1^*=0.014$. This scenario lets clear why a fluid-fluid demixing has not been observed in the 0NN-1NN and 1NN-2NN binary mixtures. A possible explanation for this fact is that, although these are nonadditive mixtures, the particle sizes are not so dissimilar as in the 0NN-2NN case to yield a demixing. Further investigations of $k$NN mixtures with larger $k$'s are necessary to confirm this. 

We have also observed an anomaly in isobaric curves of the total density of particles (versus $z_i$, with $i=0,1,2$), inside the fluid phase, which increase after passing to a minimum. Three surfaces of minimum density (SMDs) were found in the 3D parameter space, each one defined for one of the activities kept fixed. These SMDs start at different lines when we approximate the planes ($z_0,0,z_2$) and ($z_0,z_1,0$) by making $z_1\rightarrow0$ and $z_2\rightarrow0$, respectively, and ends at the spinodal surface of the fluid phase. For very large $z_2$, deep inside the metastable fluid region, these three surfaces either become quite close or collapse into a single surface. Although it is hard to decide this numerically, it is simple to figure out that this occurs because in this region one has $\rho_2 \gg \rho_0$ and $\rho_2 \gg \rho_1$, so that $\rho_T \approx 4 \rho_2$ independently of the activity being fixed. We remark that density anomalies are typically found in complex polymorphic fluids, whose modeling is usually very elaborated (as is the case, e.g., in lattice gases with directional interactions) \cite{polymorphism}. Therefore, the existence of this kind of anomaly in the simple athermal system analyzed here indicates that mixtures of hard spheres (and hard disks as well \cite{tiago15}) might be useful as a starting point to understand anomalies in more complex fluids.

\acknowledgments

We acknowledge support from CNPq, CAPES and FAPEMIG (Brazilian agencies).

\appendix
\section{Recursion relations for the ternary mixture}

The recursion relations (RRs) for the partial partition functions (ppf's) of the 0NN-1NN-2NN mixture were obtained using the method described in Sec. \ref{defmod}. For the root site in the sublattice $A_1$, they are given by
\begin{widetext}
\begin{subequations}
\begin{eqnarray}
 a'_{1,\varnothing}&=& b_{1,\varnothing} c_{1,\varnothing} d_{1,\varnothing} a_{2,\varnothing} b_{2,\varnothing} c_{2,\varnothing} d_{2,\varnothing}\{(1+z_0 B_{1,0}) (1+z_0 C_{1,0}) (1+z_0D_{1,0}) (1+z_0 A_{2,0}) (1+z_0 B_{2,0}) (1+z_0 C_{2,0})\\ \nonumber
 &&(1+z_0 D_{2,0})+ z_1(B_{1,1} + C_{1,1} + D_{1,1} + A_{2,1} + B_{2,1} + C_{2,1} + D_{2,1}) \\ \nonumber
 &+& z_1^2[B_{1,1}(D_{1,1} + A_{2,1} + B_{2,1} + C_{2,1}) + C_{1,1}(B_{2,1} + C_{2,1} + D_{2,1}) + D_{1,1}(A_{2,1} + C_{2,1} + D_{2,1}) + A_{2,1} C_{2,1} + B_{2,1} D_{2,1}] \\ \nonumber
 &+& z_1^3[B_{1,1} D_{1,1}(A_{2,1} + C_{2,1}) + B_{1,1} A_{2,1} C_{2,1} + C_{1,1} B_{2,1} D_{2,1}+ D_{1,1} A_{2,1} C_{2,1}] \\ \nonumber
 &+& z_1^4B_{1,1} D_{1,1} A_{2,1} C_{2,1} +  z_0 z_1[B_{1,1}(D_{1,0} + A_{2,0} + B_{2,0} + C_{2,0}) + C_{1,1}(B_{2,0} + C_{2,0} + D_{2,0}) \\ \nonumber
 &+& D_{1,1}(B_{1,0} + A_{2,0} + C_{2,0}+ D_{2,0}) + A_{2,1}(B_{1,0} + D_{1,0} + C_{2,0}) + B_{2,1}(B_{1,0} + C_{1,0} + D_{2,0}) \\ \nonumber
 &+& C_{2,1}(B_{1,0} + C_{1,0} + D_{1,0} + A_{2,0}) + D_{2,1}(C_{1,0} + D_{1,0} + B_{2,0})] \\ \nonumber
 &+&  z_0^2z_1[B_{1,1} D_{1,0}(A_{2,0} + B_{2,0} + C_{2,0}) + B_{1,1} A_{2,0}(B_{2,0} + C_{2,0}) + B_{1,1} B_{2,0} C_{2,0} + C_{1,1} B_{2,0}(C_{2,0} + D_{2,0}) \\ \nonumber
 &+& C_{1,1} C_{2,0} D_{2,0} + D_{1,1} B_{1,0}(A_{2,0} + C_{2,0} + D_{2,0}) + D_{1,1} A_{2,0}(C_{2,0} + D_{2,0}) + D_{1,1} C_{2,0} D_{2,0} \\ \nonumber
 &+& A_{2,1} B_{1,0}(D_{1,0} + C_{2,0}) + A_{2,1} D_{1,0} C_{2,0} + B_{2,1} B_{1,0}(C_{1,0} + D_{2,0}) + B_{2,1} C_{1,0} D_{2,0} \\ \nonumber 
 &+& C_{2,1} B_{1,0}(C_{1,0} + D_{1,0} + A_{2,0}) + C_{2,1} C_{1,0}(D_{1,0} + A_{2,0}) + C_{2,1} D_{1,0} A_{2,0} + D_{2,1} C_{1,0}(D_{1,0} + B_{2,0}) + D_{2,1} D_{1,0} B_{2,0}] \\ \nonumber
 &+&  z_0^3z_1[B_{1,1} D_{1,0} A_{2,0}(B_{2,0} + C_{2,0}) + B_{1,1} D_{1,0} B_{2,0} C_{2,0} + B_{1,1} A_{2,0} B_{2,0} C_{2,0} + C_{1,1} B_{2,0} C_{2,0} D_{2,0} \\ \nonumber
 &+& D_{1,1} B_{1,0} A_{2,0}(C_{2,0} + D_{2,0}) + D_{1,1} B_{1,0} C_{2,0} D_{2,0} + D_{1,1} A_{2,0} C_{2,0} D_{2,0} + A_{2,1} B_{1,0} D_{1,0} C_{2,0} + B_{2,1} B_{1,0} C_{1,0} D_{2,0} \\ \nonumber
 &+& C_{2,1} B_{1,0} C_{1,0}(D_{1,0} + A_{2,0})+C_{2,1} B_{1,0} D_{1,0} A_{2,0} + C_{2,1} C_{1,0} D_{1,0} A_{2,0} + D_{2,1} C_{1,0} D_{1,0} B_{2,0}] \\ \nonumber
 &+&  z_0^4z_1(B_{1,1} D_{1,0} A_{2,0} B_{2,0} C_{2,0} + D_{1,1} B_{1,0} A_{2,0} C_{2,0} D_{2,0} + C_{2,1} B_{1,0} C_{1,0} D_{1,0} A_{2,0}) \\ \nonumber 
 &+&  z_0 z_1^2[B_{1,1} D_{1,1}(A_{2,0} + C_{2,0}) + B_{1,1} A_{2,1} (D_{1,0} + C_{2,0}) + B_{1,1} C_{2,1}(D_{1,0} + A_{2,0}) + C_{1,1} B_{2,1} D_{2,0} + C_{1,1} D_{2,1} B_{2,0} \\ \nonumber
 &+& D_{1,1} A_{2,1}(B_{1,0} + C_{2,0}) + D_{1,1} C_{2,1}(B_{1,0} + A_{2,0}) + A_{2,1} C_{2,1}(B_{1,0} + D_{1,0}) + B_{2,1} D_{2,1} C_{1,0}] \\ \nonumber
 &+&  z_0^2z_1^2(B_{1,1} D_{1,1} A_{2,0} C_{2,0} + B_{1,1} A_{2,1} D_{1,0} C_{2,0} + B_{1,1} C_{2,1} D_{1,0} A_{2,0} + D_{1,1} A_{2,1} B_{1,0} C_{2,0} \\ \nonumber
 &+& D_{1,1} C_{2,1} B_{1,0} A_{2,0} + A_{2,1} C_{2,1} B_{1,0} D_{1,0}) +  z_0 z_1^3(B_{1,1} D_{1,1} A_{2,1} C_{2,0} + B_{1,1} D_{1,1} C_{2,1} A_{2,0} + D_{1,1} A_{2,1} C_{2,1} B_{1,0} \\ \nonumber
 &+& B_{1,1} A_{2,1} C_{2,1} D_{1,0}) + z_2(B_{1,2} + C_{1,2} + D_{1,2}+A_{2,2}+B_{2,2}+C_{2,2}+D_{2,2}) \\ \nonumber
 &+& z_2^2(B_{1,2} B_{2,2} + C_{1,2} C_{2,2} + D_{1,2} D_{2,2}) +  z_0 z_2(B_{1,2} B_{2,0} + C_{1,2} C_{2,0} + D_{1,2} D_{2,0} + B_{2,2} B_{1,0} + C_{2,2} C_{1,0} + D_{2,2} D_{1,0}) \\ \nonumber
 &+&  z_1 z_2(B_{1,2} B_{2,1} + C_{1,2} C_{2,1} + D_{1,2} D_{2,1} + B_{2,2} B_{1,1} + C_{2,2} C_{1,1} + D_{2,2} D_{1,1})\}
\end{eqnarray}
\begin{eqnarray}
 a'_{1,0}&=& b_{1,\varnothing} c_{1,\varnothing} d_{1,\varnothing} a_{2,\varnothing} b_{2,\varnothing} c_{2,\varnothing} d_{2,\varnothing}\{(1+z_0 B_{1,0}) (1+z_0 C_{1,0}) (1+z_0 D_{1,0}) (1+z_0 A_{2,0}) (1+z_0 B_{2,0}) (1+z_0 C_{2,0})\\ \nonumber
 &&(1+z_0 D_{2,0})+ z_1(C_{1,1} + A_{2,1} + B_{2,1} + D_{2,1}) + z_1^2(C_{1,1} B_{2,1} + C_{1,1} D_{2,1} + B_{2,1} D_{2,1}) + z_1^3C_{1,1} B_{2,1} D_{2,1} \\ \nonumber
 &+&  z_0 z_1[C_{1,1}(B_{2,0} + C_{2,0} + D_{2,0}) + A_{2,1}(B_{1,0} + D_{1,0} + C_{2,0}) + B_{2,1}(B_{1,0} + C_{1,0} + D_{2,0}) + D_{2,1}(C_{1,0} + D_{1,0} + B_{2,0})] \\ \nonumber
 &+&  z_0^2z_1[C_{1,1} B_{2,0}(C_{2,0} + D_{2,0}) + C_{1,1} C_{2,0} D_{2,0} + A_{2,1} B_{1,0}(D_{1,0} + C_{2,0}) + A_{2,1} D_{1,0} C_{2,0} \\ \nonumber
 &+& B_{2,1} B_{1,0}(C_{1,0} + D_{2,0}) + B_{2,1} C_{1,0} D_{2,0} + D_{2,1} C_{1,0} (D_{1,0} + B_{2,0}) + D_{2,1} D_{1,0} B_{2,0}] \\ \nonumber
 &+&  z_0^3z_1(C_{1,1} B_{2,0} C_{2,0} D_{2,0} + A_{2,1} B_{1,0} D_{1,0} C_{2,0} + B_{2,1} B_{1,0} C_{1,0} D_{2,0} + D_{2,1} C_{1,0} D_{1,0} B_{2,0})\\ \nonumber
 &+& z_0 z_1^2(C_{1,1} B_{2,1} D_{2,0} + C_{1,1} D_{2,1} B_{2,0} + B_{2,1} D_{2,1} C_{1,0}) + z_2 A_{2,2}\}
\end{eqnarray}
\begin{eqnarray}
 a'_{1,1}&=& b_{1,\varnothing} c_{1,\varnothing} d_{1,\varnothing} a_{2,\varnothing} b_{2,\varnothing} c_{2,\varnothing} d_{2,\varnothing}\{1 + z_0(C_{1,0} + A_{2,0} + B_{2,0} + D_{2,0}) + z_0^2[C_{1,0}(A_{2,0} + B_{2,0} + D_{2,0}) \\ \nonumber
 &+& A_{2,0}(B_{2,0} + D_{2,0}) + B_{2,0} D_{2,0}]+z_0^3[C_{1,0} A_{2,0}(B_{2,0} + D_{2,0}) + C_{1,0} B_{2,0} D_{2,0} + A_{2,0} B_{2,0} D_{2,0}] \\ \nonumber
 &+& z_0^4C_{1,0} A_{2,0} B_{2,0} D_{2,0}+z_1(C_{1,1} + A_{2,1} + B_{2,1} + D_{2,1}) + z_1^2(C_{1,1}(B_{2,1} + D_{2,1})+B_{2,1} D_{2,1}) \\ \nonumber 
 &+& z_1^3 C_{1,1} B_{2,1} D_{2,1} +  z_0 z_1[C_{1,1}(B_{2,0} + D_{2,0}) + B_{2,1}(C_{1,0} + D_{2,0}) + D_{2,1}(C_{1,0} + B_{2,0})]   \\ \nonumber
 &+& z_0^2z_1(C_{1,1} B_{2,0} D_{2,0} + B_{2,1} C_{1,0} D_{2,0} + D_{2,1} C_{1,0} B_{2,0}) \\ \nonumber
 &+&  z_0 z_1^2(C_{1,1} B_{2,1} D_{2,0} + C_{1,1} D_{2,1} B_{2,0} + B_{2,1} D_{2,1} C_{1,0}) +  z_2 A_{2,2}\}
\end{eqnarray}
\begin{eqnarray}
 a'_{1,2}&=& b_{1,\varnothing} c_{1,\varnothing} d_{1,\varnothing} a_{2,\varnothing} b_{2,\varnothing} c_{2,\varnothing} d_{2,\varnothing}\left(1 +  z_0 A_{2,0} +  z_1 A_{2,1} +  z_2 A_{2,2}\right)
\end{eqnarray}
\label{RRs1NN2NN}
\end{subequations}
\end{widetext}
where $A_{i,j}$, $B_{i,j}$, $C_{i,j}$ and $D_{i,j}$, with $i=1,2$ and $j=0,1,2$, are the ratios defined in Eq. \ref{ratio}. The RRs for the other sublattices can be obtained from cyclic permutations of the sublattice labels:  $A_1 \rightarrow B_2$, $B_2 \rightarrow C_1$, $C_1 \rightarrow D_2$, $D_2 \rightarrow A_1$, together with $C_2 \rightarrow D_1$, $D_1 \rightarrow A_2$, $A_2 \rightarrow B_1$, $B_1 \rightarrow C_2$. Following this order, we can find the RRs for the sublattices $D_2$, $C_1$ and $B_2$. Those for sublattices $A_2$, $B_1$, $C_2$ and $D_1$ can be obtained from the ones for $A_1$, $B_2$, $C_1$ and $D_2$, respectively, by exchanging all the sublattice indexes $i$ ($1 \leftrightarrow 2$). Once we have the RRs for the ppf's at hand, the ones for their ratios are obtained by simply dividing the corresponding equations, e.g., by dividing the equations \ref{RRs1NN2NN}b, \ref{RRs1NN2NN}c and \ref{RRs1NN2NN}d by \ref{RRs1NN2NN}a.

\end{document}